\documentclass{ws-ijmpd2}
\usepackage[super,compress]{cite}

\usepackage{nccmath}
\usepackage{moresize}
\usepackage{enumerate}
\usepackage[caption=false]{subfig}
\usepackage{multirow}
\usepackage{mathtools}
\usepackage{stackengine}                     

\usepackage{tikz}
\usetikzlibrary{positioning}
\usepackage{stackengine}
\usepackage{scalerel}

\stackMath

\newcommand{\be}{\begin{equation}}
\newcommand{\ee}{\end{equation}}
\newcommand{\bea}{\begin{eqnarray}}
\newcommand{\eea}{\end{eqnarray}}
\newcommand{\bse}{\begin{subequations}}
\newcommand{\ese}{\end{subequations}}
\newcommand{\bce}{\begin{center}}
\newcommand{\ece}{\end{center}}
\newcommand{\bfg}{\begin{figure}}
\newcommand{\efg}{\end{figure}}
\newcommand{\bit}{\begin{itemize}}
\newcommand{\eit}{\end{itemize}}
\newcommand{\bed}{\begin{description}}
\newcommand{\eed}{\end{description}}
\newcommand{\ben}{\begin{enumerate}}
\newcommand{\een}{\end{enumerate}}
\newcommand{\nn}{\nonumber}

\newcommand{\la}{\label}
\newcommand{\pa}{\partial}
\newcommand{\fr}{\frac}
\newcommand{\sq}{\sqrt}
\newcommand{\no}{\noindent}

\def\a  {\alpha}
\def\b  {\beta}
\def\c  {\gamma}
\def\C  {\Gamma}
\def\d  {\delta}

\def\e  {\epsilon}

\def\f  {\phi}

\def\k  {\kappa}
\def\l  {\lambda}
\def\L  {\Lambda}
\def\m  {\mu}
\def\n  {\nu}
\def\o  {\omega}
\def\O  {\Omega}

\def\r  {\rho}

\def\Th {\Theta}
\def\s  {\sigma}
\def\t  {\tau}
\def\vph {\varphi}

\def\le {\left}
\def\ri {\right}
\newcommand{\cA}{\mathcal A}

\newcommand{\cL}{\mathcal L}

\newcommand{\cQ}{\mathcal Q}

\newcommand{\cT}{\mathcal T}

\newcommand{\fw}{\mathfrak w}
\newcommand{\nab}{\nabla\!}
\newcommand{\nt}{\widetilde{\nabla}\!}
\newcommand{\at}{\widetilde{a}}
\newcommand{\ut}{\widetilde{u}}

\newcommand{\Gt}{\widetilde{G}}
\newcommand{\Rt}{\widetilde{R}}
\newcommand{\Wt}{\widetilde{W}}
\newcommand{\Ct}{\widetilde{\C}}
\newcommand{\Lt}{\widetilde{\L}}

\newcommand{\hg}{\widehat{g}}

\newcommand{\hT}{\widehat{T}}
\newcommand{\hX}{\widehat{X}}
\newcommand{\hC}{\widehat{\C}}

\newcommand{\hS}{\widehat{S}}
\newcommand{\hnab}{\widehat{\nab}}
\newcommand{\hnt}{\widehat{\nt}}
\newcommand{\cLm}{\cL^{(m)}}
\newcommand{\Sm}{S^{(m)}}

\newcommand{\hcA}{\widehat{\cA}}
\newcommand{\hcQ}{\widehat{\cQ}}
\newcommand{\hcT}{\widehat{\cT}}

\newcommand{\hlam}{\widehat{\l}}

\newcommand{\he}{\widehat{\e}}

\newcommand{\vx}{\vec{\pmb x}}

\newcommand{\keff}{\k_{\text{\scriptsize eff}}}
\newcommand{\Rteff}{\Rt^{\text{\scriptsize eff}}}

\newcommand{\Tm}{T^{(m)}}
\newcommand{\Td}{T^{(d)}}
\newcommand{\TT}{T^{(T)}}
\newcommand{\rmt}{\r^{(m)}}

\newcommand{\rct}{\r^{(c)}}

\newcommand{\rdt}{\r^{(d)}}
\newcommand{\pdt}{p^{(d)}}

\newcommand{\rTt}{\r^{(T)}}

\newcommand{\OL}{\O^{(\!\L\!)}}
\newcommand{\OLp}{\O_{_0} \!\!^{\!(\!\L\!)}}

\newcommand{\rmp}{\r^{(m)}_{_0}}

\newcommand{\rcp}{\r^{(c)}_{_0}}

\newcommand{\sw}{\mathsf w}

\newcommand{\wx}{\sw_{\!_X}}

\newcommand{\cTp}{\cT_{_0}}

\newcommand{\Hp}{H_{_0}}
\newcommand{\fp}{\f_{_0}}
\newcommand{\tp}{t_{_0}}
\newcommand{\rp}{\rho_{_0}}
\newcommand{\betp}{\b_{_0}}

\newcommand*\rfra[2]{{}^{\scriptstyle{#1}}\!\!\diagup_{\!\!\scriptstyle{#2}}}
\newcommand*\rfraa[2]{{}^{\displaystyle #1}\!\!\!\diagup_{\!\!\displaystyle #2}}

\newcommand{\bdm}{\begin{displaymath}}
\newcommand{\edm}{\end{displaymath}}

\long\def\symbolfootnote[#1]#2{\begingroup%
\def\thefootnote{\fnsymbol{footnote}}\footnote[#1]{#2}\endgroup}
\numberwithin{equation}{section}
\providecommand{\appendixname}{Appendix}

\begin{document}

\markboth{H. Ramo Chothe, Ashim Dutta and Sourav Sur}
{Cosmological Dark sector from a Mimetic-Metric-Torsion perspective}

%
\catchline{}{}{}{}{}
%

\title{\LARGE{Cosmological Dark sector from a Mimetic-Metric-Torsion perspective}}

\author{HIYANG RAMO CHOTHE}

\address{\it Department of Physics \& Astrophysics, University of Delhi\\
New Delhi - 110 007, India\\
rmo.tnm@gmail.com}

\author{Ashim Dutta}

\address{\it Department of Physics \& Astrophysics, University of Delhi\\
New Delhi - 110 007, India\\
ashim1921@gmail.com}

\author{SOURAV SUR}

\address{\it Department of Physics \& Astrophysics, University of Delhi\\
New Delhi - 110 007, India\\
sourav.sur@gmail.com; sourav@physics.du.ac.in}

\maketitle

\begin{history}
\received{Day Month Year}
\revised{Day Month Year}
\end{history}

\begin{abstract}
We generalize the basic theory of mimetic gravity by extending its purview to the general metric-compatible 
geometries that admit torsion, in addition to curvature. This essentially implies reinstating the mimetic 
principle of isolating the conformal degree of freedom of gravity in presence of torsion, by parametrizing 
both the physical metric and torsion in terms of the scalar `mimetic' field and the metric and torsion of a 
fiducial space. We assert the requisite torsion parametrization from an inspection of the fiducial space Cartan 
transformation which, together with the conformal transformation of the fiducial metric, preserve the physical 
metric and torsion. In formulating the scalar-tensor equivalent Lagrangian, we consider an explicit contact 
coupling of the mimetic field with torsion, so that the former can manifest itself geometrically as the source 
of a torsion mode, and most importantly, give rise to a viable `dark universe' picture from a mimicry of an 
evolving dust-like cosmological fluid with a non-zero pressure. A further consideration of higher derivatives 
of the mimetic field in the Lagrangian leads to physical bounds on the mimetic-torsion coupling strength, which 
we determine explicitly.  
\end{abstract}







\section{Introduction       \label{sec:intro}}

General Relativity (GR), in spite of its immense success, has been subject to quite a bit of skepticism on its adequacy 
and correctness in predictability at cosmological scales, when confronted with the challenges in dealing with the dark 
constituents of the universe, viz. {\em dark energy} (DE) and {\em dark matter} (DM)
\cite{CST-rev,AT-book,wols-ed,MCGM-ed,BCNO-rev}. 
Considerable interest has therefore been developed on the cosmological aspects of the extensions or alternatives of GR
--- collectively, the theories of {\em modified gravity} (MG)
\cite{chiba-mg,NO-mg1,NO-mg2,NO-mg3,FTT-mg,SF-mg,FT-mg,clift-mg,papa-ed,NOO-rev}.
Attention has primarily been drawn by the MG equivalent scalar-tensor formulations which, in a cosmological setting, 
give rise to scenarios of a scalar field induced DE, that interacts with matter fields (including the DM sources) under 
conformal transformations
\cite{fuj-st,frni-st,ENO-st,FTBM-st,BGP-st}.
Consequently, one gets a platform for exploring a plausible `geometric unification' of the cosmological dark sector via 
some mechanism of emulating the DM (or part thereof) using the same scalar field artefact(s) of geometry from which the 
DE emerges. Such explorations have had a fresh impetus in recent years with the emergence of the {\em mimetic gravity} 
theory 
\cite{CM-mm},
and a host of its extensions.

Mimetic gravity remarkably achieves an exact fluid description of an irrotational {\em dust}, supposedly the cold dark 
matter (CDM) in the Friedmann-Robertson-Walker (FRW) cosmological framework, while preserving the conformal symmetry of 
the {\it physical} space-time metric $g_{\m\n}$
\cite{CM-mm,CMV-mm,MV-mm,HV-mm,barv-mm,CR-mm,ramz-mm,BR-mm,SVM-mm}. 
This nonetheless follows from the diffeomorphism invariance of gravitational theories \`a la GR, which allows one to 
parametrize $g_{\m\n}$ by a fiducial metric $\hg_{\m\n}$ and a scalar field $\f$, in a non-invertible {\em disformal} 
manner in general
\cite{bek-dt,FG-dt,BT-dt,DR-dt}. 
The mimetic parametrization though is a simplified disformation, viz. $g_{\m\n} = - \k^2 \hg^{\a\b} \pa_\a \f 
\pa_\b \f \, \hg_{\m\n}$, where $\k$ denotes the Planck length scale, and the scalar `mimetic field' $\f$ is 
assumed dimensionless
\cite{CM-mm}.
Such a parametrization ensures not only the invariance of $g_{\m\n}$ under a conformal transformation of $\hg_{\m\n}$,
but also that $\f$ is left {\it non-dynamical} by the condition for the invertibility of $g_{\m\n}$, viz. $\, - \k^2 
g^{\m\n} \pa_\m \f \pa_\n \f = 1 \,$. This condition could be implemented in the theory as a constraint, using a Lagrange 
multiplier $\l$ in an equivalent mimetic action. A further equivalence of the latter with the singular Brans-Dicke (BD) 
action (in which the BD parameter $\fw = - \rfra{3\!}{2}$) eventually shows that mimetic gravity is indeed a ratification 
of GR being raised to the status of a scalar-tensor equivalent MG theory
\cite{CM-mm,CMV-mm,MV-mm,HV-mm,SVM-mm}.

Detailed studies have revealed many attractive features of the basic mimetic model of Chemseddine and Mukhanov 
(henceforth, the CM model
\cite{CM-mm})
and its various extensions, from the perspectives of both cosmology and astrophysics
\cite{RMMM-mdyn,DKSTS-mdyn,LS-mdyn,OO-mdyn,NO-mfR,NOO-mfR,MSV-mfR,OO-mfR1,OO-mfR2,CM-msing,BGY-msing,HSP-msing,MSVZ-mgal,
vag-mgal,MS-mco,MMGM-mco,AO-mco,CM-mco,nas-mco,NHB-mco,SJ-mgw,BBFLMS-mgw,LSYN-mgw,BPT-mgw,RSCV-mgw,GBKM-mgw,CRSV-mgw}.
In particular, the CM Lagrangian extended by a potential $V (\f)$ leads to the scenarios of a cosmologically evolving 
`mimetic fluid', which characterizes {\em dust}, albeit with a non-vanishing pressure\footnote{Reminiscent of the {\em dusty 
dark energy} (DDE) models  
\cite{LSV-dde,GGWC-dde,CMNO-dde}, 
that preceded the original CM paper
\cite{CM-mm}.}
\cite{CMV-mm}. 
As such, no propagating scalar perturbation mode is there to suppress the growth of structures at smaller (sub-galactic) 
scales
\cite{CR-mm,ramz-mm}, 
or to define the quantum fluctuations requisite for $\f$ to provide the seeds of the observed large scale structure of the 
universe
\cite{CMV-mm,MV-mm}. 
A reasonable supposition has therefore been to extend the CM theory further with {\it higher derivative} (HD) term(s) for 
$\f$, e.g. $(\Box \f)^2$, which lead(s) to a non-zero sound speed $c_s$ of the mimetic matter perturbations, keeping the 
background solution unaffected qualitatively\footnote{In fact, the dusty fluid with pressure is rendered {\em imperfect}, 
with or without a violation of the shift symmetry ($\f \rightarrow \f \,+$ constant) of the theory, in presence of the HD 
terms
\cite{MV-mm,ramz-mm,BR-mm}.  
}
\cite{CMV-mm,MV-mm,SVM-mm}. 
However, the HD terms in general make the theory susceptible to Ostrogradsky ghost or(and) gradient instabilities
\cite{CR-mm,ramz-mm,CP-mhd,GMF-mhd,CKOT-minst,ZSML-minst},
possible wayouts of which point to more complicated CM extensions (with explicit HD couplings with the Ricci curvature 
scalar $R$
\cite{CKOT-minst,HNK-minst,ZSML-minst,TK-minst},
degenerate higher-order scalar-tensor (DHOST) couplings  
\cite{BLN-mdhost,LMNV-mdhost},
etc.). 
Noteworthy mimetic extensions from various other considerations include the mimetic $f(R)$ gravity and its variants 
\cite{LS-mdyn,OO-mdyn,NO-mfR,NOO-mfR,MSV-mfR,OO-mfR1,OO-mfR2,MS-mnfR,MMG-mfRv,AOO-mfRv,BHHA-mfRv},
mimetic Horndeski theory
\cite{ABKM-mh,CMSVZ-mh}, 
mimetic Born-Infeld theory
\cite{BCC-mbi,CBC-mbi}, 
mimetic brane-world gravity
\cite{SN-mbw,YYZL-mbw,ZZGL-mbw},
mimetic massive gravity
\cite{CM-mmg1,CM-mmg2,MS-mmg,SVA-mmg},
and so on
\cite{MAM-mfR,MSVZ-mhor,kosh-mhd,ST-m2s,KNY-mvt}. 
It has also been shown that mimetic gravity has a close correspondence with the scalar version of the Einstein-{\AE}ther 
theory, and hence appears in the infrared limit of the projectable Ho\v{r}ava-Lifshiftz gravity 
\cite{HHSS-mmEA-1,JS-mmEA-1,HHSS-mmEA-2,JS-mmEA-2,sper-mmEA,RACMP-mhorlif}.

Nevertheless, many of the extended mimetic formulations often have implicit arbitrariness, for example in the assertion 
of the potential $V (\f)$, or in the choice of appropriate HD terms or(and) their couplings with curvature invariants, or 
in some other context. Scope therefore remains for further extensions or generalizations, particularly with the legitimacy 
of asking for a {\em proper geometrical significance} of the mimetic field, notwithstanding its role in encoding the 
conformal degree of freedom of gravity. More specifically,
\bit
\item {\em Can we have the mimetic field manifesting itself geometrically, say for e.g. acting as the source of a purely 
geometric entity?}
\eit
An answer to this may be reckoned within the option of considering such an entity to be {\em torsion}, which is the only 
space-time characteristic, other than curvature, that can be extracted from a general metric-compatible affine connection
\cite{traut,HVKN-trev,akr-tbook,SG-tbook,SS-tbook}.

Torsion's physical significance is in its indispensability in forming a classical background geometry for quantized matter 
fields with arbitrary {\it spin}, and hence as a plausible low energy manifestation of a fundamental theory of quantum 
gravity 
\cite{shap-trev,fab-thesis,blag-book,CL-extgrav}.
Indeed, an axial torsion has been argued to have its source in the string theoretic Kalb-Ramond field  
\cite{pmssg,ham},
with many implications revealed from extensive studies
\cite{ssgss-kr1,ssgas-kr,DJM-kr,skpmssgas-kr,skpmssgss-kr,skssgss-kr,ssgss-kr2,dmssgss-kr1,sssdssg-kr,AMT-kr,BC-kr,CMS-kr,
bmssssg-kr1,ssgss-kr3,dmssg-kr,dmssgss-kr2,sdadssg-kr,bmssssg-kr2,adbmssg-kr,scssg-kr1,scssg-kr2,scssg-kr3}. 
Also appealing are the physical aspects of e.g. the {\em propagating torsion} theories
\cite{HRR-proptor,CF-proptor,saa-proptor,BS-proptor,popl-proptor,BC-proptor,ND-proptor},
the {\em parity violating metric-torsion} theories
\cite{HMS-pvtor,holst-pvtor,bmssgss-pvtor,bmssssgss-pvtor,dmpmssg-pvtor,CM-pvtor,FMT-pvtor}, 
the {\em extended gravity} theories with torsion
\cite{ACCF-extgr,FC-extgr,CMT-extgr},
the {\em skewon} theory of gravity with torsion
\cite{ROH-skew,HORB-skew,Ni-skew},
the {\em scalar-torsion} theory that demonstrates Cartan gauge invariance
\cite{FRM-scaltor},
the {\em square-torsion} theory
\cite{VFS-sqtor,lu-sqtor,fab-sqtor,VCVM-sqtor},
the {\em degenerate tetrad} formalism
\cite{KS-deg1,KS-deg2,SS-deg1,KS-deg3,SS-deg2},
the non-minimal {\em metric-scalar-torsion} coupling formalism
\cite{ssasb-mst1,ssasb-mst2,ssasb-mst3},
and so on. Moreover, a lot of attention has been drawn in recent years by modified versions of the (curvature-free) 
{\em teleparallel} gravity theories 
\cite{FF-fT,BF-fT,LSB-fT,CCDDS-fT,BMT-fT,GLSW-fT,GLS-fT,IS-fT,CGSV-fT,SST-fT,KPS-fT,CCLS-fT,
BB-fT,BCFN-fT,CLPR-fT,FSGS-fT}, 
apart from the modern refinements of Poincar\'e gauge theory of gravity
\cite{YN-PGT,MGK-PGT,NSV-PGT,SNY-PGT,mink-PGT,BHN-PGT,GLT-PGT,HB-PGT,LC-PGT,NRR-PGT,obu-PGT}.

Now, given such a wide applicability of torsion, a generalized mimetic gravity formulation in presence of it seems 
really intriguing, particularly from the perspective of getting an observational support for torsion, amidst its 
miniscule experimental evidence till date
\cite{KRT-texpt,FR-texpt,BF-texpt,HOP-texpt,CCR-text,CCSZ-texpt,LP-texpt}.
In fact, there had been some expectation that the mimetic field could be identified with, or at least correlated to, 
a torsion degree of freedom
\cite{HSP-msing}.
However, only a few attempts on a mimetic-torsion formulation have so far been made, and that too in the specific
contexts of e.g. Lyra geometry, or modified teleparallelism  
\cite{MAM-mfR,MO-mfT,GZYSL-mfT}.
Whether such a formulation be substantiated within the purview of the rather general Riemann-Cartan geometry, that admits 
both curvature and torsion, is nonetheless an open issue, worth proper attention and study. It is our objective in this 
paper to do so, with an endeavour to see the role of torsion's coupling with the mimetic field in leading to viable 
cosmological scenarios. 

Let us outline the organization of this paper. After briefly reviewing the basic tenets of metric-torsion theories in 
section \ref{sec:genconf}, we focus on the conformal properties of torsion and in particular motivate a simplified form 
of the so-called Cartan transformation equation from certain standpoints. Then in section \ref{sec:torconf} we proceed 
towards formulating a mimetic-metric-torsion (MMT) theory, by (i) finding a mechanism to ensure its conformal symmetry 
at the field (metric and torsion) level, (ii) motivating and proposing the equivalent MMT Lagrangians, and (iii) working
out the corresponding field equations under certain simplifying assumptions. We particularly emphasize on formulating the 
equivalent MMT Lagrangians under the consideration of explicit contact couplings of the mimetic field $\f$ with the 
individual torsion terms\footnote{These terms are those that generally appear in the conventional metric-torsion 
Lagrangians, such as that formulated originally by Cartan and Einstein in early 1920s.}. As we show, such `MMT couplings' 
are essential for not only a geometric manifestation of $\f$ (as the source of the torsion trace mode $\cT_\m$), but also 
for retaining the dust-like characteristic of the mimetic fluid in space-times with torsion. We actually resort to a 
specific setup in which the equivalent MMT Lagrangians look elegant and simplified, with just one independent MMT coupling 
function $\b (\f)$. Ignoring for brevity any external sources for the torsion irreducible modes, we find that the system 
of MMT equations of motion look exactly similar to that for an evolving dust-like mimetic fluid with a non-zero pressure 
(resulting from the MMT coupling, instead of any ad-hoc potential term as considered in the literature
\cite{CMV-mm,SVM-mm}).
In section \ref{sec:gmmcosm} we work out the MMT cosmological equations in the FRW framework, and subsequently resort 
to a well-motivated quadratic coupling $\b (\f) \sim \f^2$, which leads to a $\L$CDM cosmological evolution, where $\L$ 
is an effective cosmological constant. We illustrate the various phases of evolution of $\b (\f)$ and the norm of $\cT_\m$ 
(which we consider as a {\em torsion parameter}), and estimate their values at the present epoch using the latest PLANCK 
results
\cite{Planck18}. 
Then in section \ref{sec:MMT-HD}, we study the outcome of incorporating in our MMT formalism a $(\Box \f)^2$ term, which
leads to a non-zero sound speed $c_s$ of the linear matter perturbations, without affecting the background ($\L$CDM) 
solution qualitatively. As such, we find that there is no change in the torsion parameter (norm of $\cT_\m$), but the 
MMT coupling $\b (\f)$ gets rescaled by a factor proportional to $(1 + 3 c_s^2)^{-1}$. Consequently, the physical limits 
of $c_s^2$ determine the bounds on $\b (\f)$ evaluated at the present epoch. We conclude with a summary and some open 
questions in section \ref{concl}. 
In the Appendix, we work out the mimetic fluid acceleration $\at_\m$ in a background space-time with torsion, and show that 
it vanishes (i.e. the fluid velocity is tangential to the time-like auto-parallel curves, just as for dust) only when 
$\cT_\m \sim \pa_\m \f$, i.e. for the torsion trace sourced by the mimetic field.

\bigskip
\no 
{\large \sl Conventions and Notations}: Throughout this paper we use (i) metric signature $\, (-,+,+,+)$, and (ii) natural 
units (the speed of light $c = 1$). We denote (i) the determinant of the metric tensor $g_{\m\n}$ by $g$, (ii) the Planck 
length parameter by $\, \k = \sq{8 \pi G_N}$ (where $G_N$ is the Newton's gravitational constant) and (iii) the values of 
parameters or functions at the present epoch by affixing a subscript `$0$'.

\section{Metric-Torsion formalism and Conformal transformation  \label{sec:genconf}}

Let us begin with the formal definition of the torsion tensor, viz.
\be \label{tor-def}
T^\a_{~\m\n} \,:=\, 2 \, \Ct^\a_{[\m\n]} \,\equiv\, \Ct^\a_{\m\n} \,-\, \Ct^\a_{\n\m} \,\,,
\ee
in the four dimensional Riemann-Cartan ($U_4$) space-time geometry, characterized by a general (asymmetric) 
affine connection 
\be \label{conn}
\Ct^\a_{\m\n} \,=\, \C^\a_{\m\n} \,+\, K^\a_{~\m\n} \,\,, \qquad \le[K^\a_{~\m\n} = \mfrac 1 2 \le(T^\a_{~\m\n} - 
T^{~\a}_{\m~\n} - T^{~\a}_{\n~\m}\ri) \ri] ,
\ee
where $\C^\a_{\m\n}$ is the four dimensional Riemannian ($R_4$) Levi-Civita connection, and the tensor $K^\a_{~\m\n}$ 
is known as `contorsion'. Eq. (\ref{conn}) is actually a re-statement of the condition of metric-compatibility 
$\, \nt_\a g_{\m\n} = 0 = \nab_\a g_{\m\n} \,$, where $\nt_\a$ is the $U_4$ covariant derivative defined in terms 
of $\Ct^\a_{\m\n}$ in the same way as the $R_4$ covariant derivative $\nab_\a$ is defined in terms of $\C^\a_{\m\n}$. 
Note also that $K_{\a\m\n} = K_{[\a\m]\n} \,$, unlike $T_{\a\m\n} = T_{\a[\m\n]}$. 

The $U_4$ analogue of the Riemannian curvature tensor $R^\a_{~\m\r\n}$ is given by 
\be \label{U4-curvtens}
\Rt^\a_{~\m\r\n} \,:=\, \pa_\r \Ct^\a_{\m\n} \,+\, \Ct^\a_{\t\r} \Ct^\t_{\m\n} \,-\, 
(\r \leftrightarrow \n) \,\,,
\ee
which (unlike $R^\a_{~\m\r\n}$) neither have any cyclicity property nor exhibit any symmetry under the interchange of 
the first and last pairs of indices. Accordingly, the $U_4$ analogue of the Ricci tensor $R_{\m\n}$, viz. $\Rt_{\m\n} 
:= \Rt^\a_{~\m\a\n}$ is not symmetric in its indices.
 
Also well-known is the following decomposition of the torsion tensor 
\be \label{tor-decom}
T^\a_{~\m\n} \,=\, \mfrac 1 3 \le(\cT_\m \, \d^\a_\n \,-\, \cT_\n \, \d^\a_\m\ri) +\, \mfrac 1 6 \, 
\e^\a_{~\m\n\r} \, \cA^\r \,+\, \cQ^\a_{~\m\n} \,\,,
\ee
in its irreducible modes, viz. 
(i) the {\em trace} vector $\, \cT_\m := T^\n_{~\m\n} \,$, 
(ii) the {\em pseudo-trace} vector $\, \cA^\m := \e^{\a\b\c\m} \,T_{\a\b\c} \,$, and 
(iii) the ({\em pseudo})-{\em tracefree} tensor $\, \cQ^\a_{~\m\n}\,$ ($= \cQ^\a_{~[\m\n]} \,$ and $\, \cQ^\a_{~\m\a}
= 0 = \e^{\a\b\c\m} \,\cQ_{\a\b\c} \,$). In terms of these modes, the $U_4$ connection, and the $U_4$ analogue of the 
Riemannian curvature scalar $R$, are given respectively by
\bea 
&& \Ct^\a_{\m\n} \,=\, \C^\a_{\m\n} \,-\, \mfrac 1 3 \le(\cT^\a g_{\m\n} -\, \cT_\m \, \d^\a_\n\ri) -\, 
\mfrac 1 {12} \e^\a_{~\m\n\r} \, \cA^\r \,+\, \cQ_{\n\m}^{~~\,\a} \,\,, 
\label{conn1} \\
&& \Rt \,=\, g^{\m\n} \Rt_{\m\n} \,=\, R \,-\, 2 \nab_\m \cT^\m \,-\, \mfrac 2 3 \cT_\m \cT^\m \,+\, 
\mfrac 1 {24} \cA_\m \cA^\m \,+\, \mfrac 1 2 \cQ_{\a\m\n} \cQ^{\a\m\n} \,\,.
\label{U4-curv}
\eea
Usually, $\Rt$ is taken to replace the standard Einstein-Hilbert Lagrangian $R$ in the conventional metric-torsion 
theories, viz. the original Einstein-Cartan formulation and many of its extensions or generalizations
\cite{HVKN-trev,akr-tbook,SG-tbook,SS-tbook,shap-trev}.

Now, in a metric-compatible space-time with torsion, a conformal transformation 
\be \label{conf}
g_{\m\n} \rightarrow g'_{\m\n} = e^{2\s} g_{\m\n} \,\,, 
\ee
is in general associated with the {\em Cartan transformation} equations
\cite{shap-trev,FRM-scaltor}: 
\bea 
&&\!\!\!\! T^\a_{~\m\n} \rightarrow\, T'^\a_{~\m\n} \,=\, T^\a_{~\m\n} \,+\, q \le(\d^\a_\m \, \pa_\n \s 
- \d^\a_\n \, \pa_\m \s \ri) \,; \qquad [q = \,\, \mbox{a constant}], 
\label{tor-tr} \\
&&\!\!\!\! \cT_\m \rightarrow \cT'_\m = \cT_\m - 3 q \, \pa_\m \s \,\,, \quad
\cA_\m \,\rightarrow \cA'_\m = \cA_\m \,\,, \quad 
\cQ^\a_{~\m\n} \rightarrow \cQ'^\a_{~\m\n} = \cQ^\a_{~\m\n} \,\,, \qquad
\label{tormode-tr}
\eea
for any given scalar function of coordinates $\s (t,\vx)$. Note that the transformed torsion tensor $T'^\a_{~\m\n}$ has 
the same decomposition (\ref{tor-decom}) in terms of the transformed modes $\cT'_\m$, $\cA'_\m$ and $\cQ'^\a_{~\m\n}$ 
(of course, with the indices being raised or lowered using the conformal metric $g'_{\m\n}$)\footnote{For convenience 
of our discussions later on, we purposefully call Eq. (\ref{conf}) the `conformal transformation' equation, and Eqs. 
(\ref{tor-tr}) and (\ref{tormode-tr}) together as the `Cartan transformation' equations. In the literature though, the 
`Cartan transformations' generally refer to the full set (\ref{conf})-(\ref{tormode-tr}) 
\cite{shap-trev,FRM-scaltor}.
}. In other words, Eqs. (\ref{tor-tr}) and (\ref{tormode-tr}) are defined in a way that $\cT'_\m$, $\cA'_\m$ and 
$\cQ'^\a_{~\m\n}$ are truely the irreducible modes of $T'^\a_{~\m\n}$
\cite{shap-trev,FRM-scaltor,einst-rel,mus-MTconf,fab-MTconf,berg-MTconf,BOS-TPconf,MM-TPconf,wri-TPconf}. 
The constant $q$ is an arbitrary numerical parameter, that typifies the conformal symmetry of the $U_4$ theory (if any). 
More specifically, the $U_4$ theory is said to be conformally symmetric in the `{\em weak}' or `{\em strong}' form
\cite{shap-trev,FRM-scaltor},
if the $U_4$ action is conformally invariant for $q = 0$ or $\neq 0$. In this paper however, we shall resort to the 
particular setting $q = 1$, which is of significance from the following points of view\footnote{The general case of 
an arbitrary $q$ is considered in our subsequent works 
\cite{RAS-MMT-1,RAS-MMT-2}.}:
\ben[(a)]
\item In general, the metric and the connection are taken as independent variables in a geometric theory of gravity. 
For any general affine connection $\Ct^\a_{\m\n}$, the corresponding curvature constructs $\Rt^\a_{~\m\r\n}$, 
$\Rt_{\m\n}$ and $\Rt$, as well as the Einstein tensor analogue $\Gt_{\m\n} = \Rt_{\m\n} - \fr 1 2 g_{\m\n} \Rt$, 
remain invariant under the so-called {\em Einstein's $\l$-transformation}
\cite{einst-rel}:
\be \label{lambda-tr}
\Ct^\a_{\m\n} \rightarrow\, \Ct'^\a_{\m\n} \,=\, \Ct^\a_{\m\n} \,+\, \d^\a_\m \pa_\n \l \,\,, 
\ee
where $\l = \l (t,\vx)$ is a arbitrary scalar function of coordinates. Now, given the specific form (\ref{conn1}) of 
$\Ct^\a_{\m\n}$, verify that the transformations (\ref{conf})-(\ref{tormode-tr}) imply
\be \label{conn-tr}
\Ct^\a_{\m\n} \rightarrow\, \Ct'^\a_{\m\n} \,=\, \Ct^\a_{\m\n} \,+\, \d^\a_\m \pa_\n \s \,+\, (q - 1) 
\le(g_{\m\n} g^{\a\b} \pa_\b \s \,-\, \d^\a_\n \pa_\m \s\ri) \,.
\ee
For $q = 1$ and $\s \equiv \l$, this corresponds to the $\l$-transformation (\ref{lambda-tr}), with one caveat though 
--- invariance of $\Rt^\a_{~\m\r\n}$, $\Rt_{\m\n}$ and $\Gt_{\m\n}$, but $\Rt \rightarrow e^{2\s} \Rt$. 
\item Recall that in the Riemannian space-time ($R_4$), one can construct using the curvature tensor 
$R_{\a\m\r\n}$, and its contractions $R_{\m\n}$ and $R$, the {\em Weyl} tensor 
\be \label{Weyl}
W_{\a\m\r\n} := R_{\a\m\r\n} - \le(R_{\a [\r} \, g_{\n] \m} - R_{\m [\r} \, g_{\n] \a}\ri)
+ \mfrac 1 6 \le(g_{\a [\r} \, g_{\n] \m} - g_{\m [\r} \, g_{\n] \a}\ri)\! R \,\,.
\ee
This has the same (anti)symmetry and cyclicity properties of $R_{\a\m\r\n}$, but is irreducible and {\em conformally 
covariant}, i.e. preserved in the mixed form $W^\a_{~\m\r\n}$ under the conformal transformation (\ref{conf}). In the 
$U_4$ space-time however, the conformal covariance of a tensor implies the invariance of the latter, in a certain 
specified form, under the full set of transformations (\ref{conf})-(\ref{tormode-tr}), which involve the arbitrary 
parameter $q$. Now, there is no straightforward way to find the $U_4$ equivalent of the Weyl tensor, that is uniquely 
defined for all values of $q$ and constructed out of the $U_4$ curvature analogues $\Rt_{\a\m\r\n}$, $\Rt_{\m\n}$ and 
$\Rt$. Nevertheless, a close inspection of the Jacobi-Bianchi identities in presence of torsion has led to the 
suggestion that the following extension of $\Rt_{\a\m\r\n}$ may be treated as the effective $U_4$ curvature tensor
\cite{fab-MTconf}:
\be \label{curv-ext}
\Rteff_{\a\m\r\n} \,:=\, \Rt_{\a\m\r\n} \,+ \le(\fr{q - 1}{3 q}\ri) \le(\cT_\a \, T_{\m\r\n} \,-\, \cT_\m \, 
T_{\a\r\n}\ri) \,.
\ee
Accordingly, the contractions $\Rteff_{\m\n} = g^{\a\r} \Rteff_{\a\m\r\n}$ and $\Rteff = g^{\m\n} \Rteff_{\m\n}$ may 
serve as the effective $U_4$ Ricci tensor and curvature scalar respectively. Note that $\Rteff_{\a\m\r\n}$ bears the  
antisymmetry properties of $\Rt_{\a\m\r\n}$, and so does the irreducible tensor
\be \label{Weyl-MT}
\Wt_{\a\m\r\n} := \Rteff_{\a\m\r\n} - \le(\Rteff_{\a [\r} \, g_{\n] \m} - \Rteff_{\m [\r} \, g_{\n] \a}\ri) + 
\mfrac 1 6 \!\le(g_{\a [\r} \, g_{\n] \m} - g_{\m [\r} \, g_{\n] \a}\ri)\! \Rteff \,, \quad
\ee
which can therefore be considered as the effective $U_4$ Weyl tensor
\cite{fab-MTconf},
since it is preserved in the mixed form $\Wt^\a_{~\m\r\n}$ under the transformations (\ref{conf})-(\ref{tormode-tr}). 
For $q = 0$ however, the above definition of $\Rteff_{\a\m\r\n}$ is invalid. So the case of the weak conformal symmetry 
cannot be addressed this way. On the other hand, the setting $q = 1$ simply means $\Rteff_{\a\m\r\n} = \Rt_{\a\m\r\n}$, 
i.e. we have a straightforward construction of the $U_4$ Weyl tensor $\Wt_{\a\m\r\n}$ in exact analogy of Eq. (\ref{Weyl}).
\een

\section{Mimetic theory and its generalization in presence of Torsion  \label{sec:torconf}}

\subsection{Conformal invariance of the physical fields \label{sec:fieldconf}}

Refer to the basic CM formalism that exploits the diffeomorphism invariance of GR in parametrizing the physical 
metric $g_{\m\n}$ by a fiducial metric $\hg_{\m\n}$ and the dimensionless `mimetic' scalar field $\f$, as
\cite{CM-mm,CMV-mm,SVM-mm}:
\be \label{mm-tr}
g_{\m\n} \,=\, \hX \, \hg_{\m\n} \,\,, \qquad \mbox{where} \qquad 
\hX \,=\, - \k^2 \, \hg^{\m\n} \, \pa_\m \f \, \pa_\n \f \,\,.
\ee
This parametrization is non-invertible, as it represents a mapping of $10 \rightarrow 11$ variables in $R_4$. Observe 
the following:
\ben[(i)]
\item The physical metric $g_{\m\n}$ remains invariant under the conformal transformation
\be \label{conf-aux}
\hg_{\m\n} \rightarrow e^{2\s} \, \hg_{\m\n} \quad \mbox{and} \quad \f \rightarrow \f \,\,, 
\ee
where $\, \s = \s (t,\vx) \,$ is a scalar function of coordinates\footnote{The conformal transformation relations 
would have been $\, \hg_{\m\n} \rightarrow e^{2\s} \hg_{\m\n}, \, \f \rightarrow e^{-\s} \f \,$, if $\f$ would have 
had a mass dimension $= 1$, instead of being dimensionless (see for e.g. Maldacena 
\cite{mal-CG}
).}.
\item The physical metric is non-singular (i.e. its inverse exists) only when
\be \label{consis}
X \,\equiv\, - \k^2 \, g^{\m\n} \, \pa_\m \f \, \pa_\n \f \,=\, 1 \,\,.
\ee
\item The Christoffel symbols corresponding to $g_{\m\n}$ and $\hg_{\m\n}$ have the relationship
\be \label{Christ-rel}
\C^\a_{\m\n} \,=\, \hC^\a_{\m\n} \,+\, \fr 1 {2 \hX} \le(\d^\a_\m \, \pa_\n \hX \,+\, \d^\a_\n \, \pa_\m \hX \,-\, 
\hg_{\m\n} \hg^{\a\b} \pa_\b \hX\ri) \,,
\ee
which implies that $\, \nab_\a g_{\m\n} = 0 = \hX \hnab_\a \hg_{\m\n} \,$, i.e. the spaces of $g_{\m\n}$ and 
$\hg_{\m\n}$ are indeed metric spaces (with the respective covariant derivatives $\nab_\a$ and $\hnab_\a$).
\een
Now, in order to extend the mimetic formalism to the metric-torsion scenario, we first need to find an appropriate 
analogous parametrization of a (conformally covariant) {\it physical} torsion field $T^\a_{~\m\n}$ in terms of the 
mimetic field $\f$ and a fiducial torsion $\hT^\a_{~\m\n}$ having its irreducible modes defined in the fiducial space 
as\footnote{Note that the parametrization (\ref{mm-tr}) of the physical metric $g_{\m\n}$ implies that the physical 
space Levi-Civita tensor is parametrized as $\e_{\a\b\c\d} = \hX^2 \he_{\a\b\c\d}$.}
\bea \label{torfid-modes}
&& \hcT_\m \,=\, \hg^{\a\n} \, \hT_{\a\m\n} \,\,, \qquad \hcA_\m \,=\, \hg_{\m\n} \, \he^{\a\b\c\n} \, \hT_{\a\b\c} 
\,\,, \nn\\
&& \hcQ^\a_{~\m\n} \,=\, \hT^\a_{~\m\n} \,-\, \mfrac 2 3 \, \hcT_{[\m} \, \d^\a_{\n]} \,-\, \mfrac 1 6 \, \hg^{\a\b} 
\, \hg^{\r\l} \, \he_{\b\m\n\r} \, \hcA_\l \,\,.
\eea
Indeed, a close inspection of (i) how the quantity $\hX = - \k^2 \hg^{\m\n} \pa_\m \f \pa_\n \f \,$ and its 
derivatives transform under the conformal transformation (\ref{conf-aux}), and (ii) the form of the associated 
Cartan transformation of the fiducial torsion, viz.
\be \label{torfid-tr}
\hT^\a_{~\m\n} \rightarrow \hT^\a_{~\m\n} + \d^\a_\m \, \pa_\n \s - \d^\a_\n \, \pa_\m \s \,\,,
\ee
reveals that one may conveniently resort to the parametrization  
\be \label{torph}
T^\a_{~\m\n} \,:=\, \hT^\a_{~\m\n} \,+\, \d^\a_{[\m} \, \pa_{\n]} (\ln \hX) \,\,,
\ee
which ensures the conformal covariance of the physical torsion, i.e. invariance in the mixed form $T^\a_{~\m\n}$ under 
the transformations (\ref{conf-aux}) and (\ref{torfid-tr}). The conformal covariance of the irreducible modes of 
$T^\a_{~\m\n}$ can consequently be verified from their relationships with the irreducible modes of $\hT^\a_{~\m\n}$: 
\be \label{torph-modes}
\cT_\m \,=\, \hcT_\m \,-\, \mfrac 3 2 \, \pa_\m (\ln \hX) \,\,, \qquad \cA_\m \,=\, \hcA_\m \,\,, \qquad \mbox{and} 
\qquad \cQ^\a_{~\m\n} \,=\, \hcQ^\a_{~\m\n} \,\,.
\ee
Moreover, using Eq. (\ref{torph}) one gets $\, \nt_\a g_{\m\n} = 0 = \hX \hnt_\a \hg_{\m\n} \,$, which verifies the 
metricity of $g_{\m\n}$ and $\hg_{\m\n}$ in the respective spaces with torsion fields $T^\a_{~\m\n}$ and $\hT^\a_{~\m\n}$, 
and covariant derivatives denoted by $\nt_\a$ and $\hnt_\a$.

\subsection{Equivalent Lagrangian formulation \label{sec:lagconf}}

Refer back again to the basic CM theory of mimetic gravity, the action for which is no different from the standard 
Einstein-Hilbert action in presence of matter fields
\be \label{mm-ac}
S =\, \Sm +\, \fr 1 {2 \k^2} \! \int \! d^4 x \sq{- g} \, R (g_{\m\n}) \,\,; \qquad \Big[\Sm = \int \! d^4 x \sq{- g} \, 
\cLm \Big],
\ee
where $\cLm$ is the matter Lagrangian. A recourse to the parametrization $g_{\m\n} \le(\hg_{\m\n}, \f\ri)$, given by 
Eq. (\ref{mm-tr}), leads to the field equations
\bea 
&& R_{\m\n} - \fr 1 2 g_{\m\n} R \,=\, \k^2 \le[\Tm_{\m\n} \,+\, \Td_{\m\n}\ri] \,, \label{mm-eq1} \\
&& \nab_\m \le[g^{\m\n} \, \Td \, \pa_\n \f\ri] =\, 0 \,\,, \label{mm-eq2}
\eea
where $\Tm_{\m\n}$ is the matter energy-momentum tensor, whose mimetic modification is
\be \label{emtens}
\Td_{\m\n} \,= \Big[R \,+\, \k^2 \Tm\Big] \pa_\m \f \, \pa_\n \f \,\,,
\ee
$\Tm \! = g^{\m\n} \Tm_{\m\n}$ and $\, \Td \!= g^{\m\n} \Td_{\m\n} \!= - \big[\k^{-2} R + \Tm\big]$ being the respective 
traces. 

Eq. (\ref{emtens}) can be recast in the form of the energy-momentum tensor due to a pressureless fluid (dubbed the 
`mimetic fluid'), viz.
\be \label{emtens1}
\Td_{\m\n} = \rdt u_\m u_\n \,\,, \quad \le[\mbox{of energy density:} \,\, \rdt = - \Td = \k^{-2} R + \Tm\ri] ,
\ee
by identifying the fluid velocity as $u_\m \equiv \k \pa_\m \f$, in analogy with k-essence cosmologies
\cite{AMS2000-kess,AMS2001-kess,MCLT-kess,schr-kess,sssd-kess},
and in accord with the relation $X = - \k^2 g^{\m\n} \pa_\m \f \pa_\n \f = 1$ [{\it cf}. Eq.(\ref{consis})]. Note 
that it is this relation which makes the velocity normalization condition $\, g^{\m\n} u_\m u_\n = - 1 \,$ hold. Moreover, 
this relation implies that the fluid acceleration
\be 
a_\m = g^{\a\n} u_\a \nab_\n u_\m = \k^2 \, \nab_\n \le(\mfrac 1 2 \, g^{\a\n} \pa_\a \f  \, \pa_\m \f\ri) = 0 \,, 
\ee
i.e. the mimetic fluid has the flow lines of its elements following the time-like geodesics
\cite{LSV-dde},
and hence behaves as the standard {\it dust} (supposedly the cold dark matter in the FRW cosmological framework  
\cite{CM-mm,CMV-mm,SVM-mm}).

It is also easy to see that the above field equations (\ref{mm-eq1}) and (\ref{mm-eq2}) can be derived from the equivalent 
action
\be \label{mm-ac1}
S \,=\, \Sm +\, \fr 1 {2 \k^2} \! \int \! d^4 x \sq{- g} \Big[R (g_{\m\n}) \,+\, \l \!\cdot\! (X - 1)\Big] \,,
\ee
where $\l$ is a scalar Lagrange multiplier field (of mass dimension $= 2$) that enforces the constraint $X = 1$
\cite{CM-mm,SVM-mm}.
A host of other equivalent mimetic actions have also been found in the literature
\cite{HV-mm}.

Let us consider, for an illustration (and for our subsequent discussions), the following Brans-Dicke (BD) action in the 
fiducial metric space:
\be \label{mm-bdac}
\hS = \fr 1 {2 \k^2} \! \int \!\! d^4 x \sq{- \hg} \le[\hX R (\hg_{\m\n}) - \fr{\fw}{\hX} \hg^{\m\n} \pa_\m \hX \pa_\n \hX 
- \hlam \!\cdot\! \le(\! \hX + \k^2 \hg^{\m\n} \pa_\m \f \, \pa_\n \f \!\ri)\! \ri] , \quad
\ee
where $\fw$ is the constant BD parameter. Note that $\hX$ is treated here just as a (dimensionless) BD scalar field, and not 
as a quantity pre-assigned to be equal to $- \k^2 \hg^{\m\n} \pa_\m \f \, \pa_\n \f$ [what we had in Eq.(\ref{mm-tr})]. In 
fact, a scalar Lagrange multiplier field $\hlam$ (of mass dimension $= 2$) is used instead, to impose $\hX = - \k^2 \hg^{\m\n} 
\pa_\m \f \, \pa_\n \f \,$ as a constraint. Moreover, $\hX$ being dimensionless, the effective gravitational coupling is 
determined as $\keff^2 = \k^2 \hX^{-1}$, for a suitable reference setting, for e.g. the stipulation that at the present epoch 
$t = \tp$, $\hX(\tp) = 1 \,$, so that $\keff^2 (\tp) = \k^2 = 8 \pi G_{_N}$. 

Under the conformal transformation
\be \label{conf1}
\hg_{\m\n} \rightarrow g_{\m\n} = \hX \hg_{\m\n} \,\,, \qquad \f \rightarrow \f \,\,, \qquad 
\hlam \rightarrow \l = \hX^{-1} \hlam \,\,,
\ee
the action (\ref{mm-bdac}) transforms to (\ref{mm-ac1}), albeit at the free theoretical level (i.e. without the matter action 
$\Sm$), if one sets $\fw = - \rfra{3\!}{\!2}$ (specific to the singular BD theory)  
\cite{HV-mm}.

Now, a generalization of the CM action, or equivalents thereof, in presence of torsion, crucially 
requires us to ponder on the following:
\ben
\item Non-minimal metric-torsion couplings with scalar fields are seemingly more favourable than the usual minimal 
couplings which lead to a well-known uniqueness problem while defining equivalent Lagrangians by integrating out 
boundary terms
\cite{shap-trev,ssasb-mst1,ssasb-mst2,ssasb-mst3}. 
Such a problem however arises only when the torsion has its trace $\cT_\m \neq 0$ and the scalar fields are dynamical,
unlike the mimetic field $\f$ in the basic CM theory. So, apparently there is no concern in generalizing this theory 
by coupling the non-dynamic $\f$ minimally with curvature and torsion. Nevertheless, provisions for a dynamical $\f$ 
is often desired in order to have extended mimetic formulations by incorporating higher derivative terms, e.g. 
$(\Box \f)^2$, in any of the equivalent actions (\ref{mm-ac}) and (\ref{mm-ac1}) describing a minimal gravitational 
coupling with $\f$
\cite{CMV-mm,SVM-mm}.
In such a situation, the generalizations of these actions in presence of a generic torsion (with $\cT_\m \neq 0$) are 
not straightforward, as the uniqueness problem becomes imminent. A better option is therefore to generalize the 
explicitly non-minimal (or, Jordan frame) action $\hS$ given by Eq. (\ref{mm-bdac}).
\item The Jordan frame action (\ref{mm-bdac}), being defined in the fiducial space with metric $\hg_{\m\n}$, its 
generalization amounts to that in presence of the fiducial torsion $\hT^\a_{~\m\n}$. Appropriate conformal and 
Cartan transformations may then lead to the Einstein frame action in the physical space, i.e. the equivalent action
generalizing (\ref{mm-ac1}), and hence (\ref{mm-ac}). Note also that the action (\ref{mm-bdac}) does not include any 
matter Lagrangian $\cLm$. In fact, a fiducial space action is anyhow not ideal for describing matter fields, since
the latter usually couple to the physical metric $g_{\m\n}$ (and torsion $T^\a_{~\m\n}$, if existent). Therefore, 
while taking matter into consideration it is preferable to first generalize the Jordan frame action (\ref{mm-bdac}), 
then get to the equivalent (Einstein frame) action in the physical space, and finally incorporate the $\cLm$ term therein.     
\item The matter Lagrangian $\cLm$ could be due to a host of external fields --- scalars, pseudo-scalars, and fields 
with different spin (e.g. fermions, vector bosons, etc.), some of which may induce the various torsion modes. However, 
the mimetic theory is not expected to get affected qualitatively if torsion exists only in presence of certain 
external matter fields. What would be really intriguing is an appropriate generalization of the CM action such that 
the mimetic field $\f$ could manifest itself geometrically, by acting as a potential source of a torsion mode.
\item The torsional generalization would have its true significance only if it leads to a dust-like fluid component, 
mimicking cold dark matter, in the standard cosmological setup. By `dust-like' we mean the fluid with not necessarily 
a vanishing pressure, but a vanishing acceleration $a_\m$, so that the fluid velocity $u_\m$ is tangential to the 
time-like geodesics
\cite{LSV-dde}.
Now, in referring to geodesics in space-times with torsion, one has to be careful since the so-called {\em affine 
geodesics} (or the {\em auto-parallel curves}), that transport their tangent vectors parallelly along themselves, do 
not in general coincide with the (Riemannian) {\em metric geodesics} which extremize the space-time interval $ds^2 = 
g_{\m\n} dx^\m dx^\n$ along themselves
\cite{SG-tbook,SS-tbook,popl-trev,WZ-trev,HDMN-trev,HO-trev,shap-trev,FRM-scaltor}. 
Therefore, in presence of torsion it is necessary to see whether the mimetic fluid velocity $u_\m$ is tangential to 
the time-like auto-parallels, and if so, then in what circumstances. In other words, under what condition (if any) 
the effective fluid acceleration $\at_\m = u^\a \nt_\a u_\m$ vanishes, thereby confirming the auto-parallel equation? 
As we show in the Appendix, one can indeed have $\at_\m = 0$, provided the torsion trace $\cT_\m \propto \pa_\m \f$, 
where $\f$ is the mimetic field. The latter therefore, must act as the source of the torsion mode $\cT_\m$ in order
that the dust-like character of the mimetic fluid is retained in space-times with torsion.
\een
Taking all these into consideration, let us resort to the following action:
\bea \label{gmm-bdac}
\hS \,=\, \fr 1 {2\k^2} \int \! d^4 x \sq{- \hg} \bigg[&& \!\!\! \hX \sum_{n=0}^4 \b_n (\f) \, P_n (\hg_{\m\n}, 
\hT^\a_{~\m\n}) \nn\\
&&- \le\{\fr{\fw (\f)}{\hX} \, \hg^{\m\n} \pa_\m \hX \, \pa_\n \hX \,+\, \eta (\f) \, \widehat \Box \hX\ri\}\bigg] \,,
\eea
where $\, \widehat \Box \equiv \hg^{\m\n} \hnab_\m \hnab_\n \,$, and $P_n$'s denote the various terms in the $U_4$
curvature scalar analogue $\Rt$ [{\it cf}. Eq. (\ref{U4-curv})] (in the fiducial space of course):
\bea \label{P-def}
P_0 = R (\hg) \,, ~&& P_1 = \hnab_\m \le(\hg^{\m\n} \hcT_\n\ri) \,, \quad P_2 = \hg^{\m\n} \hcT_\m \hcT_\n \,, \nn\\
&& P_3 = \hg^{\m\n} \hcA_\m \hcA_\n \,, \quad P_4 = \hg_{\m\a} \hg^{\n\b} \hg^{\l\c} \hcQ^\m_{~\n\l} \hcQ^\a_{~\b\c} \,\,.
\eea
This action (\ref{gmm-bdac}) has a direct bearing on the actions in the literature that have long been suggested for 
non-minimal metric-torsion couplings to a dynamical scalar field $\Phi$ with a self-interacting potential $U (\Phi)$.
Such actions are of the general form
\be \label{MST-ac}
S \,= \int \! d^4 x \sq{- \hg} \bigg[\Phi^2 \sum_{n=0}^4 \b_n P_n \,-\, \mfrac 1 2 \, \pa_\m \Phi \, \pa^\m \Phi \,-\, 
U (\Phi)\bigg] , 
\ee
with constant $\b_n$'s, and $P_n$'s the same as in Eq. (\ref{P-def}) except with the hats over the variables removed 
(see for e.g. Shapiro
\cite{shap-trev}
and references therein)\footnote{A specific combination of $\b_n$'s, in presence of matter fields, can in fact give rise 
to cosmological scenarios that exhibit a weakly dynamic dark energy
\cite{ssasb-mst1,ssasb-mst2,ssasb-mst3}.
}.
The above action (\ref{gmm-bdac}) is actually a further generalization of (\ref{MST-ac}) in the following respects:
\bit
\item There are two scalar fields involved (in the fiducial space though) --- one is $\hX$, which is a dynamical field
(an analogue of $\Phi$), and the other one is $\f$, which has no dynamical terms to begin with\footnote{Later on, of 
course, we shall consider its kinetic term of $\f$ to exist, and constrained in a way that $\f$ is interpreted as the 
mimetic field.}. 
\item The couplings $\b_n$ have in general been taken to be functions of $\f$, so as to make it act as a source of the 
trace mode of torsion, via the corresponding equation of motion (as we shall demonstrate in the next subsection).
\item The self-terms for $\hX$ have also been taken as general as possible, viz. 
\ben[(i)]
\item The term $\, \hX^{-1} \hg^{\m\n} \pa_\m \hX \, \pa_\n \hX \,$ is considered to have a variable coefficient $\fw (\f)$, 
that replaces the constant BD parameter $\fw$ in Eq. (\ref{mm-bdac}). 
\item The term $\widehat \Box \hX$ is considered to have a variable coefficient $\eta (\f)$, which of course prevents it 
from being a surface term.
\een
\eit
Note also that the non-minimal coupling of $\f$ with $R (\hg)$ (or $P_0$) in the action (\ref{gmm-bdac}) can always be 
removed by absorbing the coefficient $\b_0 (\f)\,$ in a redefinition of the field $\hX$. Appropriate redefinitions of 
the other coupling functions $\b_n (\f)$ (for $n = 1$ to $4$), as well as $\fw (\f)$ and $\eta (\f)$, can then leave
Eq. (\ref{gmm-bdac}) same as before, except with $\b_0 = 1$. Finally, we can make out a `mimetic field' interpretation
of $\f$ by including in (\ref{gmm-bdac}) a Lagrange multiplier term that enforces $\hX$ to be equal to $- \k^2 \hg^{\m\n} 
\pa_\m \f \, \pa_\n \f \,$, in the same way as in the old (torsion-free) action (\ref{mm-bdac}). The end result would be
\bea \label{gmm-bdac1}
\hS \!&=&\! \fr 1 {2 \k^2} \!\int \! d^4 x \sq{- \hg} \bigg[\hX \bigg\{R (\hg_{\m\n}) \,+\, \sum_{n=1}^4 \b_n (\f) \, 
P_n (\hg_{\m\n}, \hT^\a_{~\m\n})\bigg\} \nn\\
\!&& -\, \bigg\{\fr{\fw (\f)}{\hX} \, \hg^{\m\n} \pa_\m \hX \, \pa_\n \hX \,+\, \eta (\f) \, \widehat \Box 
\hX\bigg\}\! - \hlam \!\cdot\! \le(\hX \,+\, \k^2 \hg^{\m\n} \pa_\m \f \, \pa_\n \f\ri)\bigg] . \quad
\eea
This we propose here as the generalization of the action (\ref{mm-bdac}) in presence of torsion, i.e. our scalar-tensor
equivalent mimetic-metric-torsion (MMT) action.

Under the set of conformal and Cartan transformations, viz. 
\bea \label{Cart1}
&& \hg_{\m\n} \rightarrow g_{\m\n} = \hX \hg_{\m\n} \,, \quad \f \rightarrow \f \,, \quad \hlam \rightarrow \l = \hX^{-1} 
\hlam \,, \tag*{\mbox{[{\it cf}. Eq. (\ref{conf1})]}} \nn\\
&& \hcT_\m \rightarrow \cT_\m = \hcT_\m - \mfrac 3 2 \pa_\m (\ln \hX) \,, \quad \hcA_\m \rightarrow \cA_\m = \hcA_\m \,, 
\quad \hcQ^\a_{~\m\n} \rightarrow \cQ^\a_{~\m\n} = \hcQ^\a_{~\m\n} \,, \qquad\quad
\eea 
Eq. (\ref{gmm-bdac1}) transforms to (after eliminating surface terms):
\bea \label{gmm-ac}
S = \fr 1 {2 \k^2} \!\int \!\! d^4 x \sq{- g} \bigg[\!\!& R &\! (g_{\m\n}) \,+\, \l \!\cdot\! \le(X - 1\ri) \nn\\
&+& \b_1 (\f) \nab_\m \cT^\m + \b_2 (\f) \cT_\m \cT^\m + \b_3 (\f) \cA_\m \cA^\m + \b_4 (\f) \cQ_{\m\n\l} 
\cQ^{\m\n\l} \nn\\
&+&\! \bigg\{\pa_\m \Big[\eta (\f) - \mfrac 3 2 \b_1 (\f)\Big] - \Big[\b_1 (\f) - 3 \b_2 (\f)\Big] \cT_\m \bigg\} 
\fr{\pa^\m \hX} \hX \nn\\
&-&\! \bigg\{\fw (\f) + \mfrac 3 2 \Big[1 + \b_1 (\f) - \mfrac 3 2 \b_2 (\f)\Big]\bigg\} \fr{\pa_\m \hX \pa^\m \hX}
{\hX^2}\bigg], \qquad
\eea
where $\, X = - \k^2 g^{\m\n} \pa_\m \f \, \pa_\n \f \,$ as before [{\it cf}. Eq. (\ref{consis})], and we have 
raised and lowered the tensor indices using the transformed metric $g_{\m\n}$. 

Let us now make the following set of choices of the coupling functions: 
\bea 
&&\!\! \b_1 (\f) = 3 \b_2 (\f) = - 2 \b (\f) \,; \quad \eta (\f) = 3 \b (\f) \,; \quad \o (\f) = - \mfrac 3 2 \le[1 - 
\b (\f)\ri] ; \qquad \label{choice1} \\
&&\!\! \b_3 (\f) = \mfrac 1 {24} \, \b (\f) \,; \quad \b_4 (\f) = \mfrac 1 2 \, \b (\f) \,\,. \qquad \label{choice2}
\eea
The first set (\ref{choice1}) leaves no dynamical term for $\hX$ in the above action (\ref{gmm-ac}). The second set 
(\ref{choice2}), on the other hand, reduces (\ref{gmm-ac}) further to the particularly simplified form
\be \label{gmm-ac1}
S = \fr 1 {2 \k^2} \int \!\! d^4 x \sq{- g} \bigg[R (g_{\m\n}) \,+\, \l \!\cdot\! \le(X - 1\ri) \,+\, \b (\f) \,  
\Th (g_{\m\n}, T^\a_{~\m\n})\bigg], \qquad
\ee
where we have recalled the expression (\ref{U4-curv}) for the $U_4$ curvature scalar analogue $\Rt$, and specifically
used its torsion-dependent part:
\bea \label{T-Lagr}
\Th (g_{\m\n}, T^\a_{~\m\n}) &:=& \Rt (g_{\m\n}, T^\a_{~\m\n}) \,-\, R (g_{\m\n}) \nn\\
&=& -\, 2 \nab_\m \cT^\m \,-\, \mfrac 2 3 \cT_\m \cT^\m \,+\, \mfrac 1 {24} \cA_\m \cA^\m \,+\, \mfrac 1 2 
\cQ_{\a\m\n} \cQ^{\a\m\n} \,.
\eea
Eq. (\ref{gmm-ac1}) is the final form of our proposed MMT action (without matter fields though) in which the mimetic 
field $\f$ does not couple non-minimally with the Riemannian curvature scalar $R$, but does so with torsion. Note that 
in both the limits $\b \rightarrow 0$ and $\b \rightarrow 1$ the action (\ref{gmm-ac1}) corresponds to Eq. (\ref{mm-ac1}), 
the torsion-free mimetic action (with the matter part $\Sm$ ignored). In fact, such a correspondence is evident for any 
constant value of $\b$, whence the last term in Eq. (\ref{gmm-ac1}) becomes a surface term. The scalar-tensor action 
equivalent to (\ref{gmm-ac1}), i.e. the version of Eq. (\ref{gmm-bdac1}) one gets after making the choices (\ref{choice1}) 
and (\ref{choice2}), is given by
\bea \label{gmm-bdac2}
\hS = \fr 1 {2 \k^2} \! \int \!\! d^4 x \sq{- \hg} \bigg[\hX \!&&\! \bigg\{R (\hg_{\m\n}) \,+\, \b (\f) \, 
\Th (\hg_{\m\n}, \hT^\a_{~\m\n})\bigg\} \nn\\
&& +\, \fr{3 \le[1 - \b (\f)\ri]}{2 \hX} \, \hg^{\m\n} \pa_\m \hX \, \pa_\n \hX \,-\, 3 \b (\f) \, \widehat \Box 
\hX \nn\\ 
&& - \, \hlam \!\cdot\! \le(\hX \,+\, \k^2 \hg^{\m\n} \pa_\m \f \, \pa_\n \f\ri)\bigg] . \qquad
\eea
This of course corresponds to the torsion-free BD action (\ref{mm-bdac}), with the BD parameter $\fw = - \rfra{3\!}
{\!2}$, in the limit $\b \rightarrow 0$. However, such a correspondence is not apparent in the limit $\b \rightarrow
1$, whence $\hX$ is rendered an auxiliary field, as its self terms in the second line of Eq. (\ref{gmm-bdac2}) get
obliterated in entirety\footnote{Note that a constant times $\sq{- \hg} \, \widehat \Box \hX$ is merely a surface term.}.
Using the auxiliary equation of motion, it can indeed be verified that the action (\ref{gmm-bdac2}), albeit on-shell, 
corresponds to (\ref{mm-bdac}) with $\fw = - \rfra{3\!}{\!2}$, in the limit $\b \rightarrow 1$. Such a correspondence,
while remaining on-shell, happens for any constant value of $\b$ as well. However, in that case $\hX$ is not auxiliary, 
rather it is the torsion mode $\hcT_\m$, in the function $\Th (\hg_{\m\n}, \hT^\a_{~\m\n})$, which can be made to act 
as the requisite auxiliary field, upon eliminating a surface term.

\subsection{Field equations in presence of matter \label{sec:eomconf}}

In accord with our prior supposition that matter fields couple only to the physical metric and torsion, we shall consider 
the full MMT action to be that given by Eq. (\ref{gmm-ac1}) augmented with the matter action $\Sm$. However, we ignore 
for brevity any matter field sources for the torsion modes $\cA_\m$ and $\cQ^\a_{~\m\n}$. Such modes would therefore 
vanish by virtue of the equations of motion $\, \rfraa{\d S}{\d \cA_\m} \!= 0 \,$ and $\, \rfraa{\d S}{\d \cQ^\a_{~\m\n}} 
\!= 0 \,$. Consequently, the full MMT action is given, up to a surface term, as
\bea \label{gmm-ac2}
S = \Sm + \fr 1 {2 \k^2} \!\int \!\! d^4 x \sq{- g} \bigg[R (g_{\m\n}) &+& \l \!\cdot\! \le(X - 1\ri) \nn\\
&+& 2 \b_\f (\f) \, \cT_\m \, \pa^\m \f - \mfrac 2 3 \, \b (\f) \, \cT_\m \cT^\m\bigg] ; \quad
\eea
where $\b_\f \equiv \rfraa{d\b}{d\f} \,$. The torsion trace $\cT_\m$, thus being `auxiliary', is derived from $\f$ via the 
field equation  
\be \label{Ttr-eom}
\fr{\d S}{\d \cT_\m} = 0 \qquad \Longrightarrow \qquad \cT_\m \,=\, \fr{3 \, \b_\f (\f)}{2 \, \b (\f)} \, \pa_\m \f \,\,.
\ee
Of course, the field $\f$ has its dynamical degree of freedom taken care of by the equation of motion
\be \label{lam-eom}
\fr{\d S}{\d \l} = 0 \quad \Longrightarrow \quad X \equiv - \k^2 \pa_\m \f \, \pa^\m \f \,=\, 1 \,:\,\,\, 
\mbox{the mimetic constraint}.
\ee
This leaves us with the following two field equations:
\bea 
\fr{\d S}{\d g^{\m\n}} \!= 0 &\,\,\Longrightarrow\,\,& R_{\m\n} \!- \mfrac 1 2 g_{\m\n} R = \k^2 \Big[\Tm_{\m\n} \! 
- \! \big\{W (\f) - \l\big\} \pa_\m \f \pa_\n \f\Big] \! - \mfrac 1 2 g_{\m\n} W (\f) \,, \qquad~~ \label{met-eom}\\
\fr{\d S}{\d \f} \!= 0 &\,\,\Longrightarrow\,\,& \nab_\m \Big[\big\{W (\f) - \l\big\} \pa^\m \f\Big] =\, - \,
\fr{W_\f (\f)}{2 \k^2} \,\,, 
\label{mf-eom}
\eea
where we have defined
\be \label{W-def}
W (\f) \,:=\, - \, \fr{3 \, \b_\f^2 (\f)}{2 \, \b (\f)} \, \pa_\m \f \, \pa^\m \f \,=\, \fr{3 \, \b_\f^2 (\f)}{2 \, 
\k^2 \b (\f)} \,\,; \qquad \le[\mbox{by Eq. (\ref{lam-eom})}\ri] ,
\ee
and $\, W_\f \equiv \rfraa{d W\!}{\!d \f} \,$. The trace of Eq. (\ref{met-eom}) determines
\be \label{lam}
\l \,=\, - \, R \,+\, \k^2 \Tm \,-\, W (\f) \,\,,
\ee
which when substituted back in Eqs. (\ref{met-eom}) and (\ref{mf-eom}) yields
\bea 
&& R_{\m\n} - \fr 1 2 g_{\m\n} R \,=\, \k^2 \le[\Tm_{\m\n} \,+\, \Td_{\m\n}\ri] \,, \label{gmm-eq1} \\
&& \nab_\m \le[\le\{\k^2 \Td + 2 W (\f)\ri\} \pa^\m \f\ri] =\, - \, \fr{W_\f (\f)}{2 \k^2} \,\,, \label{gmm-eq2}
\eea
where $\, \Td_{\m\n}$, the mimetic modification of the matter energy-momentum tensor $\Tm_{\m\n}$, is now given by 
\be \label{gmm-emt}
\Td_{\m\n} \,= \le[R \,+\, \k^2 \Tm \,-\, 2 W (\f)\ri] \pa_\m \f \, \pa_\n \f  \,-\, g_{\m\n} \, \fr{W (\f)}
{2 \k^2} \,\,,
\ee
with trace $\, \Td = - \le[\k^{-2} R + \Tm\ri]$ remaining the same as that we had in the torsion-free scenario (see 
subsection \ref{sec:lagconf}).

It is easy to verify that the above field equations (\ref{gmm-eq1}) and (\ref{gmm-eq2}) would also follow from 
the action
\be \label{gmm-ac3}
S \,=\, \Sm \,+ \fr 1 {2 \k^2} \int \!\! d^4 x \sq{- g} \Big[R (g_{\m\n}) \,+\, \l \!\cdot\! \le(X - 1\ri) -\,
W (\f) \Big] \,.
\ee
Therefore, the function $W (\f)$ can naturally be interpreted as an effective {\em mimetic potential}, the same as 
that one considers in the torsion-free mimetic extensions 
\cite{CMV-mm,SVM-mm}. 
Moreover, the expression (\ref{gmm-emt}) for $\Td_{\m\n}$ is of the form of a perfect fluid, viz.
\be \label{gmm-fluid}
\Td_{\m\n} \,= \Big[\rdt \,+\, \pdt\Big] u_\m \, u_\n  \,+\, g_{\m\n} \, \pdt \,\,,
\ee
with $u_\m \equiv \k \pa_\m \f \,$, and the energy density and pressure identified respectively as
\be \label{gmm-denspr}
\rdt \,=\, \fr R {\k^2} \,+\, \Tm \,-\, \fr{3 W (\f)}{2 \k^2} \qquad \mbox{and} \qquad \pdt \,=\, -\, \fr{W (\f)}
{2 \k^2} \,\,.
\ee

\section{Cosmological aspects of the Mimetic-Metric-Torsion formalism \label{sec:gmmcosm}}

Let us resort to the standard spatially flat FRW cosmological framework, viz. that of the four-dimensional 
space-time foliated to three-dimensional maximally symmetric hypersurfaces of constant cosmic time $t$. In 
order have such a foliation and hence the FRW metric structure preserved in presence of torsion, one requires 
the torsion modes to be constrained as follows 
\cite{ssasb-tcons}:
\bit
\item Only the component $\cT_0$ of $\cT_\m$ can in general exist.
\item Only the component $\cA_0$ of $\cA_\m$ can in general exist.
\item All the components of $Q^\a_{~\m\n}$ have to vanish identically.
\eit
These constraints of course hold for our MMT formalism here, since we have $\cA_\m = 0 = Q^\a_{~\m\n}$, and Eq. 
(\ref{Ttr-eom}) implies that among the components of $\cT_\m$, only $\cT_0 \neq 0$ in general, once we have the 
mimetic field $\f = \f (t)$ in the standard cosmological setting. In fact, in the FRW space-time the mimetic 
constraint $X = 1$ [{\it cf.} Eq. (\ref{lam-eom})] implies $\dot\f \equiv \rfraa{d\f}{dt} = \rfraa 1 \k \,$. 
Therefore, without loss of generality, we can identify $\, \k \f \equiv t \,$, the cosmic time. 

Now, since we are primarily interested in the late-time cosmological evolution, we take into consideration in this
paper only the dominant constituent of the external matter, presumably the non-relativistic (baryonic) {\em dust}, 
characterized by zero pressure and with energy density  
\be \label{matdens}
\rmt (t) =\, \rmp a^{-3} (t)\,\,,
\ee
where $a (t)$ denotes the cosmological scale factor, and $\rmp$ is the value of $\rmt$ at the present epoch $t = 
\tp$ (or $a = 1$). Accordingly, the trace of the matter energy-momentum tensor $\, \Tm = - \rmt\,$. So, we have 
the critical (or total) density of the universe given by
\bea \label{crit}
\r \,=\, \rmt +\, \rdt \,=\, \fr R {\k^2} \,+\, 3 \, \pdt \,\,, \qquad \mbox{where} \qquad \pdt \,=\, -\, \fr W {2 
\k^2} \,\,.
\eea
The Friedmann equation and the energy-momentum conservation relation are as usual the ones derived from the field 
equations (\ref{gmm-eq1}) and (\ref{gmm-eq2}):
\bea 
&& H^2 (a) \,=\, \fr{\k^2} 3 \, \r (a) \,=\, \fr{\k^2} 3 \le[\rmp \, a^{-3} +\, \rdt (a)\ri] \,, \label{Fried} \\
&& \dot{\rho}^{(d)} (a) \,=\, \fr 1 \k \, \fr{d \rdt (a)}{d\f} \,=\, - \, 3 \, H (a) \le[\rdt (a) \,+\, 
\pdt (a)\ri] \,, \label{dark-consv}
\eea
where $\, H = \rfraa{\dot{a}} a \,$ is the Hubble parameter (the overhead dot $\{\cdot\}$ denoting $\,\rfraa d {\! dt}$).

\subsection{$\L$CDM evolution from a specific mimetic-torsion coupling \label{sec:mt-lcdm}}

Depending on the choices of the coupling function $\b (\f)$, that determines the effective potential $W (\f)$, we 
can have various emergent cosmological scenarios. Now, while extending the basic CM theory of mimetic gravity by a 
supplementary potential term $V (\f)$, a host of forms of the latter has been proposed in the literature
\cite{CMV-mm,SVM-mm}.
The objective has been to build up viable cosmological models of not only the evolving dark sector, but also inflation, 
bouncing universe, etc. Such forms of the potential may be legitimate for our MMT cosmological setup as well, since 
under the identification $W (\f) \equiv 2 \k^2 V (\f)$, the expressions for $\rdt$ and $\pdt$ in Eq. (\ref{gmm-denspr}) 
match precisely with those derived from the CM Lagrangian augmented with $V (\f)$
\cite{CMV-mm}.
Qualitatively however, one must keep reservations on vehemently identifying $V (\f)$ and $W (\f)$, since the choice of 
a particular form of the latter requires a rather deep-rooted theoretical understanding, i.e. a proper motivation for 
the corresponding form of the coupling function $\b (\f)$. Consider for instance, the choice $V (\f) \sim \f^{-2}$ which 
has been argued to give rise to a quintessence-like evolution in the usual (torsion-free) mimetic cosmology
\cite{CMV-mm}.
We can of course assume the same functional form of the potential in the MMT scenario, i.e. $W (\f) \sim \f^{-2}$, leading 
to a quintessence-like evolution, but by solving Eq. (\ref{W-def}) we find the requisite coupling $\b (\f) \sim (\ln \f)^2$, 
which is nonetheless weird and has no obvious motivation. 

Another important issue is the feasibility of a MMT cosmological configuration that allows for a crossing of the so-called 
{\em phantom barrier}, i.e. the equation of state parameter $\wx$ for the effective dark energy part of the mimetic dark 
sector transiting from a value $> -1$ to a value $< -1$ at a low redshift. In general indeed, mimetic gravity allows for 
such a crossing without involving any ghost (or phantom) field. One only needs to content with the mimetic field and resort 
to an appropriate form of its potential $V (\f)$ (or $W (\f)$ in the MMT context), which may however be multi-valued (or 
even discontinuous). Finding such an appropriate form of the potential is of course not an easy task. Nevertheless, in our 
subsequent paper 
\cite{RAS-MMT-1},
we happen to come across a soliton-like kink potential $\, W (\f) \sim A - B \f^{- 2}\,$, which is found to lead to phantom 
crossing at a low redshift, for specific settings of the (presumably positive-definite) constants $A$ and $B$. If we resort 
to this form of $W (\f)$ for the MMT formalism we have developed in this paper, then by solving Eq. (\ref{W-def}) we get 
$\, \b (\f) \sim B f^{-2} (\f) \le[1 + f (\f) \tan^{-1} f (\f)\ri]^2 \,$, where $\, f (\f) = \sq{B \le(A \f^2 - B\ri)^{-1}}$. 
This once again is cumbersome and cannot be reasoned from any simple viewpoint\footnote{We have no such problem though 
in our subsequent work
\cite{RAS-MMT-1},
in which we extend the MMT formalism (developed in this paper) in a way that the effective cosmological evolution is driven
by both the torsion vector modes $\cT_\m$ and $\cA_\m$. Specifically, such an extended MMT formalism deals with two 
mimetic-torsion coupling functions, viz. $\b (\f)$ and $\c (\f)$, simple power-law forms of which effectively make $\, W (\f) 
\sim A - B \f^{- 2}\,$, with preassigned values of $A$ and $B$.}.

Given such motivational difficulties, we in this paper restrict our attention to the MMT cosmological evolution typified by 
a quadratic coupling function, viz. $\b (\f) \sim \f^2$. The reason for this is two-fold: 
\bed
\item Firstly, it could be well-motivated from the point of view of ameliorating the non-uniqueness problem one encounters in 
choosing equivalent Lagrangians for scalar coupled metric-torsion theories at the minimal level (see the point (1) in the 
discussion in subsection \ref{sec:lagconf}). In fact, it is this very reason why the $\f^2$ coupling has been the trademark 
of the metric-scalar-torsion theories studied comprehensively in the literature for a long time
\cite{shap-trev,ssasb-mst1,ssasb-mst2,ssasb-mst3}.
\item Secondly, it is easy to see from Eq. (\ref{W-def}) that this quadratic coupling leads to a constant value of the effective 
potential $W$, which we may identify as a cosmological constant $\L$ (modulo a dimensional factor): 
\be \label{lcdm-W}
W \equiv\, 2 \k^2 \L \,=\, \mbox{constant} \,\,. 
\ee
Assuming then the entire dark sector to be due to the mimetic fluid, one has its energy density and pressure given respectively as
\be \label{lcdm-denspr}
\rdt \,=\, \rct \,+\, \L \qquad \mbox{and} \qquad \pdt \,=\, - \L \,\,,
\ee
under the stipulation (in Eqs. (\ref{gmm-denspr})) 
\be \label{lcdm-R}
R \,=\, \k^2 \le[\rmt \,+\, \rct \,+\, 4 \L \ri] \,,  
\ee
where $\, \rct (t) = \rcp a^{-3} (t)\,$ denotes the energy density due to the cold dark matter (CDM) in the form of a pressureless 
dust and with value $\rcp$ at the present epoch $t = \tp$. Hence, the quadratic coupling regenerates the cosmological evolution
described by the base $\L$CDM model, which is the most favoured model of the universe observationally.
\eed
Eq. (\ref{W-def}), in accord with the identification (\ref{lcdm-W}), enables us to conveniently express the quadratic coupling as
\be \label{beta-lcdm}
\b (\f) \,=\, \betp \le(\fr{\f}{\fp}\ri)^2 \,, \qquad \mbox{where} \qquad \fp = \fr{\tp}{\k} \,\,, \qquad
\betp = \b|_{\tp} = \fr{\k^2 \L \tp^2} 3 \,\,.
\ee
Moreover, for $\L$CDM the Friedmann equation (\ref{Fried}) reduces to the form
\be \label{lcdm-Fried}
H^2 (a) \,=\, \Hp^2 \, \fr{\OLp}{\OL (a)} \,\,,
\ee
where $\, \Hp = H|_{\tp}\,$ is the Hubble constant, and 
\be \label{l-dens}
\OL (a) \,=\, \fr \L {\r (a)} \,=\, \fr{\OLp \, a^3}{1 \,-\, \OLp \le(1 - a^3\ri)} \,,
\ee
is the $\L$-density parameter, with value $\OLp$ at the present epoch $t = \tp$ (or $a \!=\! 1$).
Solving the conservation equation (\ref{dark-consv}) one then obtains
\be \label{phi-sol}
\f (a) \,=\, \fp \le[\fr{\tanh^{-1} \sq{\OL (a)}}{\tanh^{-1} \sq{\OLp}}\ri] \,. 
\ee
Conversely, since $\, \rfraa \f {\!\fp} = \rfraa t \tp \,$, we get the well-known $\L$CDM cosmological solution
\be \label{a-sol}
a (t) \,= \le[\fr{1 - \OLp}{\OLp} \, \sinh^2 \le(\fr t \tp \, \tanh^{-1} \sq{\OLp}\ri)\ri]^{\rfra 1 3} \,.
\ee
It also follows from the above equations that
\be \label{lcdm-age}
\tp \,=\, \fr 2 {3 \Hp} \, \fr{\tanh^{-1} \sq{\OLp}}{\sq{\OLp}} \,\,,
\ee
which is of course the standard expression for the present age of the $\L$CDM universe.

\subsection{The evolving torsion parameter and mimetic-torsion coupling \label{sec:para-lcdm}}

Let us have a closer look in to the factors responsible for the $\L$CDM evolution, viz. 
\ben[(i)] 
\item the trace mode $\cT_\m$ of torsion, and 
\item the mimetic coupling $\b (\f)$ which induces it. 
\een
In general, in metric-torsion theories it is often convenient to treat the norms of the torsion modes, viz.
\be \label{t-norms}
\cT = \sq{- \cT_\m \cT^\m} \,\,, \quad \cA = \sq{- \cA_\m \cA^\m} \,\,, \quad \mbox{and} \quad
\cQ = \sq{- \cQ_{\m\n\s} \cQ^{\m\n\s}} \,\,,
\ee
as the {\em torsion parameters} that quantify the contribution of torsion in observable phenomena
\cite{ssasb-mst1}.
However, since in the present context $\cA_\m = 0 = \cQ^\a_{~\m\n}$, we have only one torsion parameter ---
the norm (or length) $\cT$ of the torsion trace vector $\cT_\m$. 

For a physical interpretation of $\cT$, let us refer back to the action (\ref{gmm-ac2}) in subsection \ref{sec:eomconf}. 
If the coupling $\b (\f)$ had not been there (i.e. if we would have taken the limit $\b (\f) \rightarrow 1$), then the 
mode $\cT_\m$ would not have been sourced by the mimetic field $\f$, but may have been sourced by say, some matter field 
described by $\Sm$, the matter action in (\ref{gmm-ac2}). In such a case, we would have had the {\em bare} energy-momentum 
tensor due to torsion
\be \label{Tbare-emt}
\TT_{\m\n} \,=\, - \fr 2 {\sq{-g}} \, \fr \d {\d g^{\m\n}} \bigg(- \fr{\sq{-g}}{3 \k^2} \, \cT_\a \cT^\a\bigg) =
\, \fr 2 {3 \k^2} \bigg(\cT_\m \cT_\n \,-\, \fr 1 2 g_{\m\n} \cT_\a \cT^\a\bigg) \,,  
\ee
and consequently in the FRW framework, the {\em bare} torsional energy density 
\be \label{Tbare-dens}
\rTt \,=\, - \, g^{00} \, \TT_{00} \,=\, \fr{\cT^2}{3 \k^2} \,\,.
\ee
For a non-trivial coupling $\b (\f)$ however, $\cT_\m$ is sourced by the mimetic field $\f$ via the equation of motion 
(\ref{Ttr-eom}). The norm of $\cT_\m$ (i.e. the torsion parameter) is then
\be \label{T-norm}
\cT \,\equiv\, \sq{- \, \cT_\m \, \cT^\m} \,=\, \fr{3 \, |\b_\f (\f)|}{2 \k \b (\f)} \,\,.
\ee
Working out again the energy-momentum tensor due to the terms involving torsion in the action (\ref{gmm-ac2}), one finds 
that $\rTt$, which is of course no longer the bare torsion density, gets modulated by the factor $\b (\f)$ to give the net 
contribution of torsion to the total energy density $\r$. In fact, as is evident from the field equations (\ref{met-eom}) 
and (\ref{mf-eom}), the mimetic constraint (\ref{lam-eom}) reduces this torsional contribution to $\rfraa{W\! (\f)}{\!2 
\k^2}$, where $W (\f)$ is the effective potential defined by Eq. (\ref{W-def}). Indeed, from the equations (\ref{W-def}), 
(\ref{Tbare-dens}) and (\ref{T-norm}) we get
\be \label{eff-pot}
W (\f) \,=\, \mfrac 2 3 \b(\f) \, \cT^2 \,=\, 2 \k^2 \, \b (\f) \, \rTt \,\,,
\ee
which explicitly shows that $W (\f)$ results from $\rTt$ modulated by $\b (\f)$.

Now, the $\L$CDM solution is a consequence of having $W = 2 \k^2 \L$, a constant. It would therefore be interesting to see 
how this constant results, when both the constituting factors, viz. the torsion parameter $\cT$ and the coupling $\b$, 
evolve with time (or the scale factor $a$). By Eq. (\ref{phi-sol}), the given expressions (\ref{T-norm}) and (\ref{beta-lcdm}), 
for $\cT$ and $\b$ respectively, reduce to  
\bea 
\cT (a) \!&=&\! \cTp \bigg[\fr{\tanh^{-1} \sq{\OLp}}{\tanh^{-1} \sq{\OL (a)}}\bigg] \,; \qquad \cTp = \cT|_{\tp} = \fr 
3 \tp \,\,, \label{T-evol} \\
\b (a) \!&=&\! \betp \bigg[\fr{\tanh^{-1} \sq{\OL (a)}}{\tanh^{-1} \sq{\OLp}}\bigg]^2 \,; \qquad \betp = \fr{\k^2 \L 
\tp^2} 3 = \le(\Hp \tp\ri)^2 \OLp \,\,. \label{beta-evol}
\eea
Using Eq. ({\ref{lcdm-age}), and the best fit $\OLp$ ($= 0.6889$) obtained from the combined analysis of Planck 
TT,TE,EE+lowE, Lensing and Baryon Acoustic Oscillation (BAO) data for the base $\L$CDM model
\cite{Planck18},
we estimate
\bea 
&& \cTp \,=\, \fr{9 \, \Hp \sq{\OLp}}{2 \, \tanh^{-1} \sq{\OLp}} \,=\, 3.1436 \,\,\, \mbox{in units of $\Hp$} \,\,, 
\label{T0-estm} \\
&& \betp \,=\, \fr 4 9 \bigg(\! \tanh^{-1} \sq{\OLp}\bigg)^2 =\, 0.6274 \,\,. \label{beta0-estm}
\eea
Moreover, using Eqs. (\ref{Fried}), (\ref{lcdm-W}), (\ref{lcdm-Fried}) and (\ref{eff-pot}) we have the $\L$-density parameter 
\be \la{l-dens1}
\OL \,=\, \fr \L \r \,= \le(\fr \cT {3 H}\ri)^2 \b \,\,.
\ee
Refer now to the left panel of Fig. \ref{Mmt-fig1}, which shows how the rationalized parameters $\rfraa{\cT\!} {\!\cTp}$ and 
$\rfraa{\b\!}{\!\betp}$ evolve with time, or rather with the redshift $z = a^{-1} - 1$, from a certain epoch in a fairly 
distant past, say $z = 4$, to the extreme future limit $z \rightarrow -1$. 
%
\begin{figure}[htb]
\captionsetup[subfloat]{labelformat=empty}
\centering
\subfloat[{\small $\, z  \longrightarrow$}]{\includegraphics[scale=0.55]{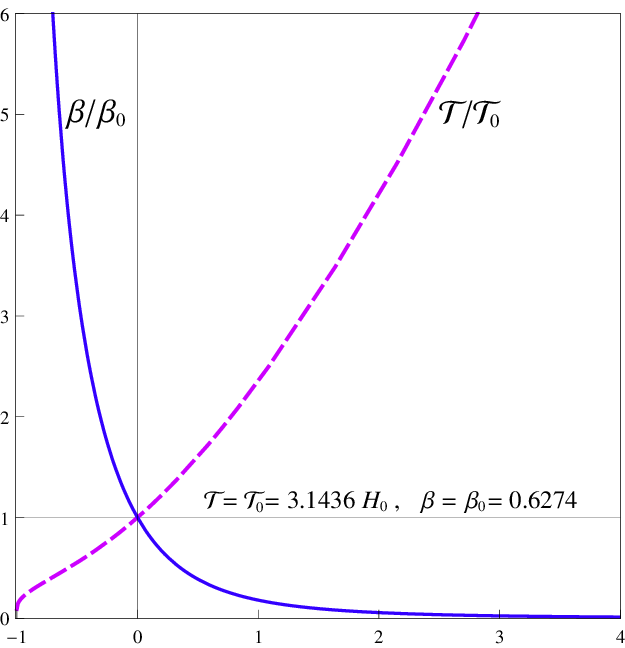}}\quad
\subfloat[{\small $\, z  \longrightarrow$}]{\includegraphics[scale=0.57]{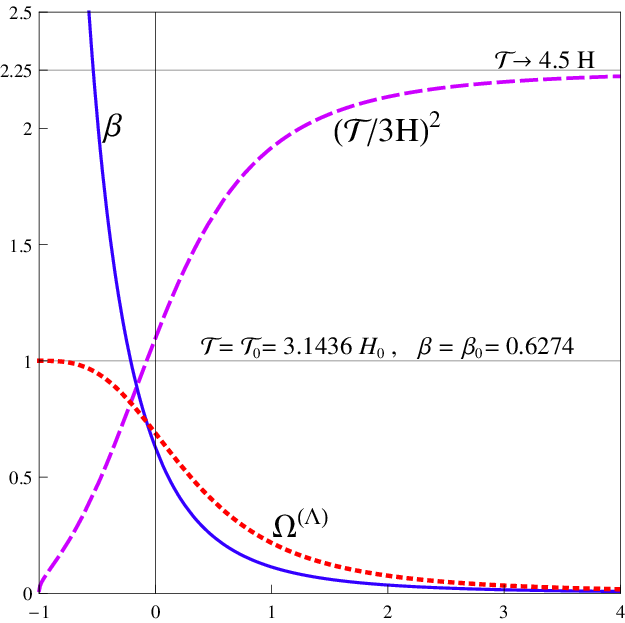}}
\caption{\footnotesize [Left panel] Evolution of the rationalized parameters $\rfraa{\cT\!}{\!\cTp}$ and $\rfraa{\b\!}
{\!\betp}$, over a redshift range $z = 4$ (in the past) to $z = - 1$ (extreme future). [Right panel] Evolution of 
$\le(\!\rfraa{\cT\!}{3 H}\!\ri)^{\!\!2}$, $\b$ and their resultant $\OL$ over the same redshift range. All the plots 
are specific to $\L$CDM, and the vertical line at $z = 0$ in either panel corresponds to the present epoch.}
\label{Mmt-fig1}
\end{figure}
%
As we see, with decreasing $z$, $\rfraa{\cT\!}{\!\cTp}$ falls off steeply in the past, until beginning to slow down a bit 
near $z \simeq 1$. It continues to decrease with $z$ at slower and slower rates till the present epoch (indicated by the 
vertical line at $z = 0$), when $\cT = \cTp$, and in the future ($z < 0$), and finally begins to diminish quickly to zero 
asymptotically as $z \rightarrow -1$. On the other hand, as $z$ decreases, $\rfraa{\b\!}{\!\betp}$ increases very slowly 
in the distant past and grows more and more rapidly afterwards from $z \simeq 1$ to the present epoch $z = 0$, when $\b = 
\betp$, and in the future, till blowing up in the limit $z \rightarrow -1$. 

The right panel of Fig. \ref{Mmt-fig1} depicts the evolution of the factors in Eq. (\ref{l-dens1}), viz. $\Big(\rfraa{\cT}
{3 H}\Big)^2$ and $\b$, and their resultant $\OL$, with $z$ in the same range $(-1,4]$. From the distant past to a fairly 
distant future, the evolution of $\OL$ is almost similar to that of $\b$, i.e. a slow increase with decreasing $z$ for $z 
\gtrsim 1$, followed by a progressively rapid growth as $z$ reduces further and further. Deep in the future ($z \lesssim 
- 0.5$) however, the growth of $\OL$ slows down considerably, unlike that of $\b$, and in the asymptotic limit $\OL 
\rightarrow 1$. 
%
\begin{figure}[htb]
\captionsetup[subfloat]{labelformat=empty}
\centering
\subfloat[{\small $\, \OL  \longrightarrow$}]{\includegraphics[scale=0.55]{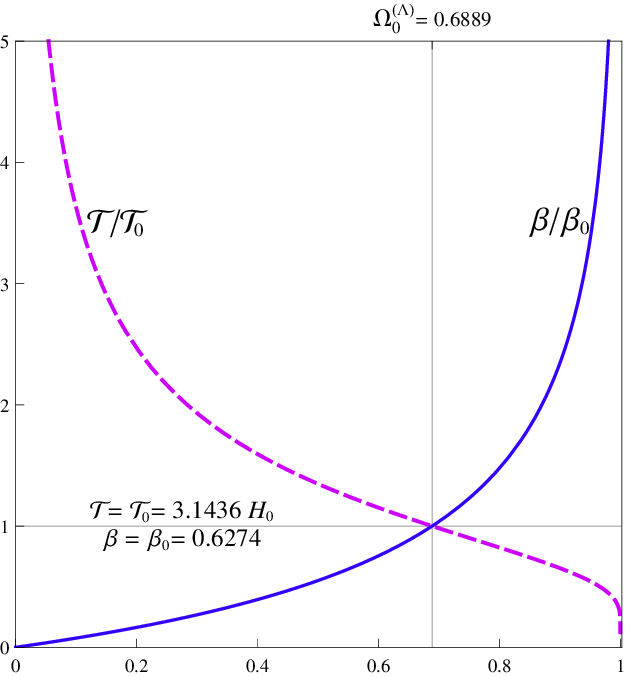}}\quad
\subfloat[{\small $\, \OL  \longrightarrow$}]{\includegraphics[scale=0.57]{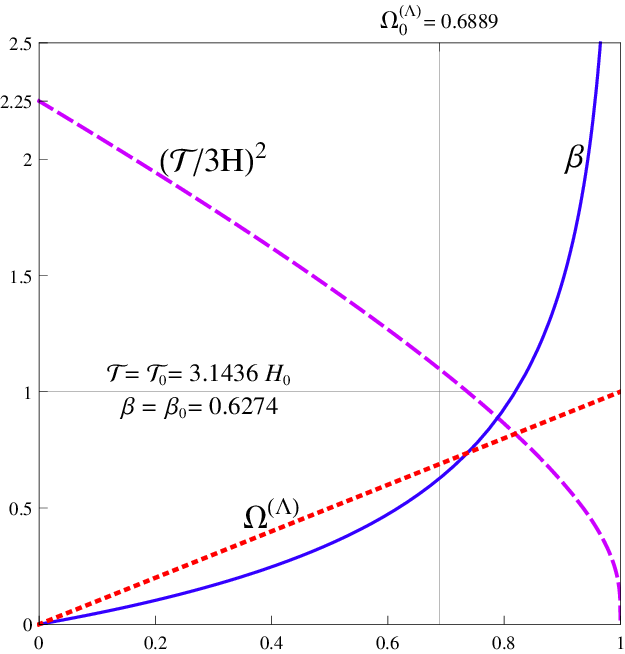}}
\caption{\footnotesize [Left panel] Variation of the rationalized parameters $\rfraa{\cT\!}{\!\cTp}$ and $\rfraa{\b\!}
{\!\betp}$ with the $\L$-density parameter $\OL$ in the entire physical range $0 \leq \OL \leq 1$. [Right panel] 
Variation of $\le(\!\rfraa{\cT\!}{3 H}\!\ri)^{\!\!2}$ and $\b$, as well as their resultant $\OL$, with $\OL$ in the 
same range. All the plots are specific to $\L$CDM, and the vertical line at $\OL = \OLp = 0.6889$ in either panel 
corresponds to the present epoch.}
\label{Mmt-fig2}
\end{figure}
%
This is of course what one expects in a $\L$CDM universe, however the crucial point to note here is that the factors 
$\cT$ and $\b$, which are responsible for the $\L$CDM solution, have their course of evolution changing drastically at 
late times. At early times ($z \gtrsim 4$), the torsion parameter $\cT$ just scales as $4.5$ times the Hubble rate $H$, 
whereas both the mimetic-torsion coupling $\b$ and the $\L$-density parameter $\OL$ are negligible. It therefore seems 
reasonable to limit one's attention up to a redshift $z \simeq 4$ in the past, beyond which $\cT$ and $\b$ are not 
expected to have any further change in their course of evolution depicted in Fig. \ref{Mmt-fig1}. 

Nevertheless, for completeness, let us resort to Fig. \ref{Mmt-fig2}, which shows the parametric variation with $\OL$,
over the entire physically acceptable domain $0 \leq \OL \leq 1 \,$ (not just over a limited range of redshift $\, z 
\in (-1,4]$ as in Fig. \ref{Mmt-fig1}). The left panel of Fig. \ref{Mmt-fig2} shows the plots for $\rfraa{\cT\!}{\!\cTp}$ 
and $\rfraa{\b\!}{\!\betp}$, whereas the right panel shows the plots for the factors $\le(\rfraa{\cT}{3 H}\ri)^2$ and 
$\b$, and their resultant $\OL$. The vertical line in either panel corresponds to the present epoch at which $\OL = \OLp 
= 0.6889$ (Planck~TT,TE,EE+lowE+Lensing+BAO best fit). As $\OL$ increases from zero, $\rfraa{\cT\!}{\!\cTp}$ falls off 
sharply, but begins to slow down considerably near $\OL \simeq 0.3$. It continues that way till at $\OL \simeq 0.9$ its 
rate of decrease starts to become faster again. $\rfraa{\cT\!}{\!H}$ however, decreases steadily with increasing $\OL$ 
until diminishing to zero rapidly as $\OL \rightarrow 1$. On the other hand, $\rfraa{\b\!}{\!\betp}$ (and in fact $\b$ 
as well) increases slowly as $\OL$ increases from zero, until growing at progressively faster rates for $\OL \gtrsim 
0.3$. Thus, once again we see that $\cT$ and $\b$ have their course of evolution changing drastically in the range $0.3 
\lesssim \OL \lesssim 0.9$, which corresponds to the late-time evolution ($-0.4 \lesssim z \lesssim 0.7 \,$, not 
stretching to very distant future though). 

On the whole, the above Eqs. (\ref{beta-evol}) -- (\ref{l-dens1}), as well as the parametric evolution shown in Figs. 
\ref{Mmt-fig1} and \ref{Mmt-fig2}, lead us to infer the following:
\bit
\item Despite getting weakened further and further as time progresses, torsion has the strength of its coupling with the 
mimetic field increasing at such a rate that the evolution of the universe is effectively driven by a constant potential 
$W \propto \b (\f) \cT^2 \,$, in addition to the pressureless (baryonic plus mimetic) matter, which we have already had 
in the torsion-free scenario. 
\item In particular, a dominant dark sector at late times is perceivable mainly from the rapid growth of the coupling 
$\b (\f)$ of the mimetic field $\f$ with torsion, although the latter decays to progressively smaller amounts with time
(which could be a plausible reason behind the miniscule experimental evidence of direct effects of torsion searched 
extensively till date
\cite{KRT-texpt,FR-texpt,BF-texpt,HOP-texpt,CCR-text,CCSZ-texpt,LP-texpt}).     
\eit

\section{Higher derivative extension and mimetic-torsion coupling limits \label{sec:MMT-HD}}

So far we have only had the cosmological evolution of a mimetic fluid with a non-zero pressure, resulting from an effective 
potential $W (\f)$. However, the presence of the potential makes with no characteristic difference of the fluid with pure 
dust, since the sound speed $c_s$ of the mimetic matter perturbations remains vanishing. The reason is that the mimetic 
constraint (\ref{lam-eom}) prevents the scalar field $\f$ to exercise its usual dynamical degree of freedom to admit any 
wave-like or oscillatory solution that one would usually expect in the linear perturbation analysis. As such, there arises 
the well-known problem of defining quantum fluctuations to $\f$ from which the large scale structures of the universe can 
originate
\cite{CMV-mm,CR-mm,MV-mm,ramz-mm,SVM-mm}.
Also, for the universe constituted by the dust-like mimetic matter, it is not possible to suppress the over-abundance of 
small scale structures
\cite{MV-mm,ramz-mm,BR-mm}.  
A possible way to ameliorate these issues is by the including higher derivative (HD) terms for $\f$ in the mimetic action. 
The most commonly used such term has been $(\Box \f)^2$, which does not qualitatively alter the cosmological solution one 
has in absence of it. This term could effectively be obtained from the action of another scalar field coupled to the 
mimetic field $\f$ in a quite non-trivial way
\cite{CMV-mm}.
Moreover, a deeper theoretical reasoning for the inclusion of such a term comes from an insight to the correspondence of 
the mimetic theory with the projectable Horava-Lifshitz gravity, the infrared limit of which has precisely the $(\Box \f)^2$ 
term appearing in the corresponding action 
\cite{RACMP-mhorlif}.

Let us see the cosmological consequences of incorporating the $(\Box \f)^2$ term in our MMT action (\ref{gmm-ac2}), or more 
conveniently, the equivalent action (\ref{gmm-ac3})\footnote{It is worth pointing out here that the $(\Box \f)^2$ term may 
actually be an inherent part of the MMT formalism itself, since as shown explicitly in our subsequent papers
\cite{RAS-MMT-1,RAS-MMT-2}, 
such a term could be induced by the axial mode $\cA_\m$ of torsion, and couplings thereof with the mimetic field $\f$.}. The 
latter takes the form
\be \label{gmm-ac4}
S \,=\, \Sm \,+ \int \!\! d^4 x \sq{- g} \le[\fr{R (g_{\m\n}) \,+\, \l \!\cdot\! \le(X - 1\ri) -\, W (\f)}{2 \k^2} \,+\, 
\fr \c 2 \!\le(\Box \f\ri)^2 \ri] \,,
\ee
where $\c$ is a dimensionless coupling factor, which we presume here to be a constant\footnote{Note that apart from the 
technical simplification, the constancy of the coefficient of the $(\Box \f)^2$ term implies that latter does not disturb 
the scale invariance of the mimetic theory.}. As is well-known, the $(\Box \f)^2$ augmentation only leads to a rescaling of 
the effective potential by a dimensionless constant in the FRW space-time
\cite{CMV-mm}.
So, the action (\ref{gmm-ac4}) could be recast in its earlier form (\ref{gmm-ac3}), after only a replacement:
\be \label{W-sclg}
W (\f) \,\longrightarrow\, \Wt (\f) \,= \le(1 \,-\, \fr{3 \c} 2\ri)^{-1} W (\f) \,\,.
\ee
Consequently, the cosmological equations and the $\L$CDM solution we have had earlier (in subsection \ref{sec:mt-lcdm}) 
remain unaltered, except that the $\L$-density parameter is now given by $\, \OL = \rfraa \Lt \r \,$, where
\be \label{L-sclg}
\Lt \,= \le(1 \,-\, \fr{3 \c} 2\ri)^{-1} \L \,\,,
\ee
is the new (rescaled) cosmological constant. From Eq. (\ref{T-evol}) it is evident that the torsion parameter $\cT$ 
remains unchanged as well. However, the MMT coupling factor $\b$, given by Eq. (\ref{beta-evol}), gets
rescaled since its strength $\betp$ at the present epoch $\tp$ is determined by the old value of the cosmological 
constant, i.e. $\L = \rfraa{W\!}{\!2 \k^2} \,$. In terms of the new value $\Lt$,
\be \label{beta0-sclg}
\betp \,= \le(1 \,-\, \fr{3 \c} 2\ri) \fr{\k^2 \, \Lt \, \tp^2} 3 \,= \le(1 \,-\, \fr{3 \c} 2\ri) \le(\Hp \tp\ri)^2 
\OLp \,\,,
\ee
where $\Hp \tp$ is as given by Eq. (\ref{lcdm-age}), and we now have $\, \OLp = \OL|_{\tp} \!= \rfraa \Lt \rp \,$, with 
$\, \rp = \r|_{\tp} \!= \rfraa{3 \Hp^2}{\!\!\k^2} \,$ being the value of the critical density $\r$ at the present epoch 
$\tp$.

Now, the striking outcome of the $(\Box \f)^2$ augmentation is the non-zero sound speed of the mimetic matter perturbations
\cite{CMV-mm}
\be \label{sndspeed}
c_s \,=\, \sq{\fr \c {2 \,-\, 3 \c}} \,\,.
\ee
Therefore, using Eq. (\ref{lcdm-age}) we get from Eq. (\ref{beta0-sclg}),
\be \label{beta0-sclg-1}
\betp \,= \fr 4 {9 \le(1 + 3 c_s^2\ri)} \le(\tanh^{-1} \sq{\OLp}\ri)^2 \,\,.
\ee
Demanding that the squared sound speed should be within its physical limits, viz. $\, 0 \leq c_s^2 \leq 1 \,$, we have 
$\, 0 \leq \c \leq \rfra 1 2$, and hence  
\be \label{beta0-bounds}
\fr 1 9 \le(\tanh^{-1} \sq{\OLp}\ri)^2 \leq\, \betp \,\leq \fr 4 9 \le(\tanh^{-1} \sq{\OLp}\ri)^2 \,\,,
\ee
where the lower(upper) bounds are those that correspond to the upper(lower) bounds of $c_s^2$. Given the Planck TT,TE,EE+lowE+ 
Lensing+BAO best fit $\, \OLp = 0.6889 \,$
\cite{Planck18},
one has $\, \betp \in \le[0.1568, 0.6274\ri]$.

\section{Conclusions \label{concl}}

We have thus generalized the basic formalism of the mimetic gravity theory
\cite{CM-mm}
for the metric-compatible ($U_4$) geometries characteristed by both curvature and torsion. Essentially, we have looked to 
extend the mimetic principle of isolating the conformal degree of freedom of gravity in the $U_4$ space-time, via the 
parametrization of not just the physical metric $g_{\m\n}$ by a fiducial metric $\hg_{\m\n}$ and the mimetic scalar field 
$\f$, but the physical torsion $T^\a_{~\m\n}$ by a fiducial torsion $\hT^\a_{~\m\n}$ and $\f$ as well. Form a close 
inspection of (i) how the quantity $\hX = - \k^2 \hg^{\m\n}\pa_\m \f \pa_\n \f$ and its derivatives transform under the 
conformal transformation of $\hg_{\m\n}$, and (ii) the form of the associated Cartan transformation equation for 
$\hT^\a_{~\m\n}$, we have been able to assert the latter and how it parametrizes $T^\a_{~\m\n}$. Of course, the foremost 
criterion for such an assertion has been to make sure that both the physical fields $g_{\m\n}$ and $T^\a_{~\m\n}$ are 
preserved under the conformal and Cartan transformations of $\hg_{\m\n}$ and $\hT^\a_{~\m\n}$ respectively. The Cartan 
transformation however, has been considered to be that of a specific class, which provides some technical simplifications 
discussed in section \ref{sec:genconf} of this paper.

While formulating the mimetic-metric-torsion (MMT) Lagrangian we have considered explicit contact couplings 
of the individual torsion terms with the mimetic field $\f$, following certain criteria which are enlisted in subsection 
\ref{sec:lagconf}. The outcome of such couplings is two-fold: (i) they implicate  $\cT_\m \propto \pa_\m \f$, meaning that 
the mimetic field $\f$ manifests itself geometrically as the source of the torsion trace mode $\cT_\m \,$, and (ii) they
give rise to an effective potential $W (\f)$, so that the dust-like mimetic fluid can have a non-zero pressure. If, for a
specific choice of $W (\f)$, the pressure is negative and large enough, a dark universe picture could be perceived in the 
standard FRW cosmological framework, just as in the phenomenological extensions of the original CM model of mimetic gravity
\cite{CMV-mm,SVM-mm}.

The dark universe interpretation however depends on whether in the first place the mimetic fluid retains its dust-like 
nature in space-times with torsion. So, there arises the question: {\em can we still have zero mimetic fluid acceleration, 
i.e. the mimetic fluid velocity tangential to the time-like auto-parallels in presence of torsion?} The answer is: {\em yes},
and as we have shown rigorously in the Appendix, the fluid acceleration indeed vanishes whenever the torsion trace $\cT_\m 
\propto \pa_\m \f$. One of the criteria mentioned above has therefore been to set up the effective Lagrangian in such a way 
that one gets $\cT_\m \propto \pa_\m \f$ naturally as an equation of motion. This is actually what prompted us to consider 
the explicit MMT coupling terms in our proposed action (\ref{gmm-bdac}) in subsection \ref{sec:lagconf}. Note also that such 
an action is a scalar-tensor action, involving non-minimal gravitational and MMT couplings with a Brans-Dicke 
scalar field $\hX$, as well as the self terms for $\hX$ and a Lagrangian multiplier term that enforces the constraint $\hX = 
- \k^2 \hg^{\m\n} \pa_\m \f \pa_\n \f$. We have chosen this action following another major criterion which is motivated from 
the point of view of avoiding a well-known uniqueness problem with the usual minimal couplings of scalar fields with torsion. 
Thereafter, performing the conformal and Cartan transformations we have worked out the equivalent minimally coupled (Einstein 
frame) MMT action (\ref{gmm-ac}), and for a certain restrictive choice of the coupling functions, the rather 
simplified action (\ref{gmm-ac1}). Augmenting the latter straightaway with the external matter action, we have then 
determined the corresponding field equations, which show that the mimetic fluid acquires an effective pressure $\pdt \propto 
- W (\f)$. For simplicity however, we have ignored any matter field source for the other torsion modes, viz. the pseudo-trace 
vector $\cA_\m$ and the (pseudo-)tracefree tensor $\cQ^\a_{~\m\n}$.

A detailed study of the cosmological aspects of our MMT formalism in section \ref{sec:gmmcosm} has revealed a plausible $\L$CDM 
evolution in the FRW framework, if the MMT coupling function is set to be $\b (\f) \sim \f^2$. In fact, for such a setting the 
potential $W$ is a constant, which could be identified as an effective cosmological constant $\L$, modulo a factor $(2 \k)^2$. 
The mimetic field $\f$, on the other hand, has its usual identification with the cosmic time $t$ in the FRW space-time. Using 
the solution for $\f$ (or $t$) in terms of the scale factor $a$ for $\L$CDM, we have made a careful examination of how the 
induced torsion field, quantified by the norm $\cT$ of the trace vector $\cT_\m$, and the quadratic coupling $\b (\f) \sim \f^2$, 
evolve with time (or, rather with the decreasing redshift $z = a^{-1} - 1$). It is found that the torsion parameter $\cT$ is 
weakened more and more, whereas the coupling $\b$ gets enhanced increasingly rapidly as time progresses. The combined effect of 
the two is of course the constant potential $W$. Moreover, both $\cT$ and $\b$ have their course of evolution changing drastically 
at late times. Therefore, despite a significant suppression in the amount of torsion, we have a late time dominance of the dark 
sector constituents, viz. the cosmological constant $\L$ (or $W$) and the dust-like mimetic matter which acts as the cold dark 
matter (CDM). The values of $\cT$ and $\b$ at the present epoch, viz. the quantities $\cTp$ and $\betp$, have been estimated using 
the Planck 2018 results, combined with weak lensing and BAO data 
\cite{Planck18}.
In particular, the Planck TT,TE,EE+lowE+Lensing+BAO best fit value of the $\L$-density parameter at the present epoch, $\OLp$, 
has been used to get the plots showing the evolution of $\cT$ and $\b$ in Figs. \ref{Mmt-fig1} and \ref{Mmt-fig2} in section 
\ref{sec:gmmcosm}. This is reasonable since the $1 \s$ error bars on the estimated parameters are very low for such a 
combination of refined data.  

The coupling $\b (\f) \sim \f^2$, which leads to the $\L$CDM solution, is reminiscent of that in the metric-scalar-torsion (MST)
theories which involve dynamical scalar fields coupled non-minimally to curvature and torsion
\cite{shap-trev,ssasb-mst1,ssasb-mst2,ssasb-mst3}. 
In the mimetic theory however, the scalar field $\f$ cannot exercise its dynamical degree of freedom because of the mimetic 
constraint. Nevertheless, it is often desired to have an implicitly dynamical $\f$ in an extended mimetic gravity formulation 
that incorporates higher derivative term(s), e.g. $(\Box \f)^2$, in the Lagrangian
\cite{CMV-mm}.
In section \ref{sec:MMT-HD}, we have considered such a $(\Box \f)^2$ term to go with the coupling $\b (\f) \sim \f^2$, in our 
MMT formalism, which is therefore liable to correspond to the MST theories, if the mimetic constraint is enforced in the latter 
using a Lagrange multiplier.

Now, the significance of the $(\Box \f)^2$ augmentation of the mimetic Lagrangian is in making the sound speed $c_s$ of the 
mimetic matter perturbations non-vanishing
\cite{CMV-mm}.
This is essential for defining the quantum fluctuations to the mimetic field $\f$ in the usual way, so that in the standard 
cosmological setting $\f$ can provide the seeds of the observed large scale structure of the universe
\cite{CMV-mm,SVM-mm}.
The background (unperturbed) cosmological solutions are however not altered qualitatively by the $(\Box \f)^2$ term, since the 
latter only leads to a rescaling of the effective potential of the mimetic field by a constant factor, that depends on $c_s^2$, 
in the FRW space-time. Accordingly, the $\L$CDM solution, that has resulted from the $\b (\f) \sim \f^2$ coupling in our MMT 
formalism, remains as the background solution, albeit with a $c_s^2$-dependent rescaling of the cosmological constant $\L$. 
Taking the observed density parameter $\OL$ as that due to the rescaled $\L$, we find the torsion parameter $\cT$ unchanged, 
but the MMT coupling strength at the present epoch, $\betp$, rescaled by a factor $(1 + 3 c_s^2)^{-1}$. The upper and lower 
bounds on $\betp$ are then determined from the demand that $c_s^2$ should be within its physical limits $0$ and $1$ respectively.

On the whole, no matter how self-sufficient our MMT formalism may seem, there remains some open questions, such as the following:
\bit
\item What about the Ostrogradsky ghost and gradient instabilities that usually arise in the mimetic gravity theory whenever 
one incorporates the $(\Box \f)^2$ and other higher derivative terms in order to have a propagating scalar perturbation mode
\cite{CKOT-minst,HNK-minst,ZSML-minst,TK-minst}? 
Can we resolve the problem of having such instabilites via explicit non-minimal couplings of some primordial field with 
curvature, torsion and the mimetic field $\f$, or by assuming one of the torsion modes to be propagating and coupled 
non-minimally to $\f$?   
\item Would it anyhow worth extending our MMT formalism by considering the torsion modes $\cA_\m$ or(and) $\cQ^\a_{~\m\n}$ to 
be existent? Under what conditions such extensions would lead to a $\L$CDM cosmological dark sector, or a small deviation from 
that? 
\item How would the torsional reparametrization and consequently the equivalent Lagrangian formulation be affected if we resort to 
the most general form of the Cartan transformation equations? What is actually the status of the Cartan gauge invariance when the 
MMT formalism is compared to the (dynamical) scalar-torsion theories in the literature 
\cite{shap-trev,FRM-scaltor}?
\item What about the first order (Palatini) reformulation of mimetic theory, for a non-minimal gravitational coupling 
with the mimetic field that would induce torsion (or, rather the {\em contorsion} part of the spin connection)?
\eit
Some of these are being addressed in a few of our subsequent works
\cite{RAS-MMT-1,RAS-MMT-2},
which we hope to report soon. 

\section*{Acknowledgement}
The authors acknowledge useful discussions with Mohit Sharma. The work of AD is supported by the University Grants Commission 
(UGC), Government of India.

\section*{Appendix: Mimetic fluid acceleration in space-times with torsion \label{App-MFaccl}}

Let us resort to the general action, used to describe a {\em dusty dark energy} (DDE) fluid, with non-vanishing pressure
\cite{LSV-dde}:
\be \label{dde-ac}
S \,=\, \int \!\! d^4 x \sq{- g} \le[\fr{R (g_{\m\n}) \,+\, \l \!\cdot\! \le(X - V (\f)\ri)}{2 \k^2} \,+\, P (\f,X) \ri] , 
\ee
where $\f$ is a dimensionless scalar field with a potential $V (\f)$, $\, X = - \k^2 \pa_\m \f \, \pa^\m \f \,$, $\l$ is 
a scalar Lagrange multiplier field (of mass dimension $= 2$) and $P (\f,X)$ is a non-canonical kinetic term for $\f$ 
(similar to what one finds in the k-essence theories
\cite{AMS2000-kess,AMS2001-kess,MCLT-kess,schr-kess,sssd-kess}).
For $V (\f) = 1$ and $P (\f,X) = 0$, one recovers the action (\ref{mm-ac1}) for mimetic gravity (in absence of external 
matter fields though). 

While the Lagrange multiplier $\l$ in the action (\ref{dde-ac}) implements the constraint
\be \label{dde-constr}
X \,\equiv\, - \k^2 \pa_\m \f \, \pa^\m \f \,=\, V (\f) \,\,,
\ee
the conserved total energy-momentum tensor, due to the fields $\f$ and $\l$, is 
\be \label{dde-emt}
T_{\m\n} \,= \le(2 \k^2 \fr{\pa P}{\pa X} \,+\, \l\ri) \pa_\m \f \, \pa_\n \f \,+\, g_{\m\n} P \,\,.
\ee
This is of the form of a perfect fluid, viz.
\be \label{dde-fluid}
T_{\m\n} \,= \le(\r \,+\, p\ri) u_\m \, u_\n  \,+\, g_{\m\n} \, p \,\,,
\ee
once we identify, in analogy with the k-essence models
\cite{AMS2000-kess,AMS2001-kess,MCLT-kess,schr-kess,sssd-kess}),
the fluid velocity as
\be \label{dde-vel}
u_\m \,=\, \fr{\k \, \pa_\m \f}{\sq{X}} \,\,, \quad \le[\mbox{so that} \quad u_\m u^\m = - 1\ri] \,,
\ee
and the fluid energy density and pressure respectively as
\be \label{dde-denspr}
\r \,=\, 2 X \fr{\pa P}{\pa X} \,-\, P \,+\, \fr{\l X}{\k^2} \qquad \mbox{and} \qquad p \,=\, P \,\,.
\ee
The fluid acceleration is then
\be \label{dde-accl}
a_\m \,:=\, u^\n \, \nab_\n \, u_\m \,=\, \k^2 \le(\nab_\m \pa_\n \f \,-\, \fr{\pa_\m \f \, \pa_\n X}{2 X}\ri) 
\fr{\pa^\n \f} X \,=\, 0 \,\,,
\ee
by virtue of the constraint (\ref{dde-constr}). So, the fluid velocity is tangential to the time-like geodesics in the
Riemannian space-time, i.e. the fluid emulates the dust, even with a non-zero pressure. Hence the name {\em dusty fluid}.

Now, the geodesics of actual geometrical significance are the {\em affine geodesics} (or the {\em auto-parallel curves}), 
which parallel transport their tangent vectors along themselves. Strictly speaking therefore, the dusty fluid is the one 
for which the fluid velocity is tangential to the time-like auto-parallels. This is ensured by the vanishing of the
fluid acceleration $a_\m$ [{\it cf}. Eq. (\ref{dde-accl})] in the Riemannian space-time, in which the auto-parallels 
coincide with the {\em metric geodesics} that extremize the space-time interval $ds^2 = g_{\m\n} dx^\m dx^\n$ between 
two neighbouring points. In non-Riemannian space-times however, such a coincidence is not guaranteed
\cite{SG-tbook,SS-tbook,popl-trev,WZ-trev,HDMN-trev,HO-trev,shap-trev,FRM-scaltor,ssasb-tcons}. 
Hence, the auto-parallel equation that should correspond to the vanishing of the dusty fluid acceleration, is not ensured 
because the latter in general differs from its Riemannian expression (\ref{dde-accl}). To be more specific, consider the 
situation in Riemann-Cartan ($U_4$) space-time that involves a non-zero torsion $T^\a_{~\m\n}$. Even if we resort to the 
minimal coupling prescription, i.e. simply replace the Riemannian covariant derivatives $\nab_\m$ everywhere by the 
Riemann-Cartan ones, $\nt_\m$, the Riemannian expressions for the velocity and acceleration, $u_\m$ and $a_\m$, would in 
general get altered. In fact, the generalization of $u_\m$ gives back its Riemannian expression (\ref{dde-vel}):
\be \label{RC-vel}
u_\m \,\longrightarrow\, \ut_\m \,=\, \fr{\k \, \nt_\m \f}{\sq{X}} \,=\, \fr{\k \, \pa_\m \f}{\sq{X}} \,=\, u_\m \,\,,
\ee
but $a_\m$ generalizes to a different form:
\be \label{RC-accl}
a_\m \,\longrightarrow\, \at_\m \,=\, \ut^\n \, \nt_\n \, \ut_\m \,=\, \k^2 \le(\nt_\m \pa_\n \f \,-\, \fr{\pa_\m \f 
\, \pa_\n X}{2 X} \,- \le[\nt_\m , \nt_\n\ri]\! \f\ri) \fr{\pa^\n \f} X\,\,.
\ee
Comparing this with Eq. (\ref{dde-accl}) we see that $a_\m = 0$ does not necessarily mean $\at_\m = 0$, because of the 
presence of certain extra terms. If these extra terms vanish as well, then only the dusty fluid will have its properties 
retained in the $U_4$ space-time. Let us figure out the criterion for the extra terms to vanish:

From the definition of $U_4$ covariant derivative of the gradient of a scalar
\be \label{RC-cov}
\nt_\m \pa_\n \f \,:=\, \nab_\m \pa_\n \f \,-\, K^\a_{~\n\m} \pa_\a \f \,\,,
\ee
where $K^\a_{~\n\m} = \mfrac 1 2 \le(T^\a_{~\n\m} - T^{~\a}_{\n~~\m} - T^{~\a}_{\m~~\n}\ri)$ is the contorsion tensor,
and the well-known commutation relation
\cite{shap-trev,ssasb-tcons}
\be \label{RC-comm}
\le[\nt_\m , \nt_\n\ri]\! \f \,=\, T^\a_{~\m\n} \, \pa_\a \f \,\,,
\ee
it is easy to show that Eq. (\ref{RC-accl}) reduces to
\be \label{RC-accl1}
\at_\m \,=\, a_\m \,-\, \fr{\k^2 \, T^\a_{~\m\n} \, \pa_\a \f \, \pa^\n \f} X \,\,,
\ee
with $a_\m$ having the expression (\ref{dde-accl}). In terms of the irreducible torsion modes, given by Eq. 
(\ref{tor-decom}), we have
\be \label{RC-accl2}
\at_\m \,=\, a_\m \,+\, \mfrac 1 3 \bigg(\cT_\m \,+\, \fr{\k^2 \pa_\m \f \, \pa^\n \f \, \cT_\n} X\bigg) - \, 
\cQ_{\a\n\m} \, \pa^\a \f \, \pa^\n \f \,\,.
\ee
The last term vanishes identically by the properties $\, \cQ^\a_{~[\m\n]} = 0 = \e^{\a\r\m\n} \cQ_{\r\m\n}$. 
Moreover, since $X = - \k^2 \pa_\m \f \pa^\m \f \,$, we infer that the second term would vanish as well (leaving $\at_\m 
= a_\m = 0$), if $\cT_\m \propto \pa_\m \f$. Note also that the relationship (\ref{RC-accl2}) between $\at_\m$ and $a_\m$ 
is irrespective of the enforcement of the constraint (\ref{dde-constr}), which has of course been responsible for the 
original result $a_\m = 0$ [{\it cf}. Eq. (\ref{dde-accl})].



\begin{thebibliography}{99}

\bibitem{CST-rev}
E. J. Copeland, M. Sami and S. Tsujikawa, {\em Dynamics of dark energy}, Int. J. Mod. Phys. 
{\bf D 15} (2006) 1753, [hep-th/0603057].

\bibitem{AT-book}
L. Amendola and S. Tsujikawa, {\em Dark Energy: Theory and Observations}, Cambridge University 
Press, United Kingdom (2010).

\bibitem{wols-ed}
G. Wolschin, {\em Lectures on Cosmology: Accelerated expansion of the Universe}, Springer, Berlin,
Heidelberg (2010).

\bibitem{MCGM-ed}
S. Matarrese, M. Colpi, V. Gorini and U. Moschella, {\em Dark Matter and Dark Energy: A Challenge 
for Modern Cosmology}, Springer, The Netherlands (2011).

\bibitem{BCNO-rev}
K. Bamba, S. Capozziello, S. Nojiri and S.D. Odintsov, {\em Dark energy cosmology: the equivalent
description via different theoretical models and cosmography tests}, Astrophys. Space Sci. {\bf 342} 
(2012) 155, [1205.3421].


\bibitem{chiba-mg}
T. Chiba, {\em $1/R$ gravity and scalar-tensor gravity}, Phys. Lett. {\bf B 575} (2003) 1,
[astro-ph/0307338].

\bibitem{NO-mg1}
S. Nojiri and S. D. Odintsov, {\em Modified Gauss-Bonnet theory as gravitational alternative for 
dark energy}, Phys. Lett. {\bf B 631} (2005) 1, [hep-th/0508049].

\bibitem{NO-mg2}
S. Nojiri and S. D. Odintsov, {\em Modified $f(R)$ gravity consistent with realistic cosmology: 
From matter dominated epoch to dark energy universe}, Phys. Rev. {\bf D 74} (2006) 086005,
[hep-th/0608008].

\bibitem{NO-mg3}
S. Nojiri and S. D. Odintsov, {\em Introduction to modified gravity and gravitational alternative 
for dark energy}, Int. J. Geom. Methods Mod. Phys. {\bf 04} (2007) 115, [hep-th/0601213].

\bibitem{FTT-mg}
S. Fay, R. Tavakol and S. Tsujikawa, {\em $f(R)$ gravity theories in Palatini formalism:
Cosmological dynamics and observational constraints}, Phys. Rev. {\bf D 75} (2007) 063509,
[astro-ph/0701479].

\bibitem{SF-mg}
T. P. Sotiriou and V. Faraoni, {\em $f(R)$ Theories of Gravity}, Rev. Mod. Phys. {\bf 82} (2010)
451, [0805.1726].

\bibitem{FT-mg}
A. De Felice and S. Tsujikawa, {\em $f(R)$ theories}, Living Rev. Rel. {\bf 13} (2010) 3, 
[arXiv:1002.4928].

\bibitem{clift-mg}
T. Clifton, P. G. Ferreira, A. Padilla and C. Skordis, {\em Modified Gravity and Cosmology}, 
Phys. Rept. {\bf 513} (2012) 1, [1106.2476].

\bibitem{papa-ed}
E. Papantonopoulos, {\em Modifications of Einstein's Theory of Gravity at Large Distances}, Lecture 
Notes in Physics, Springer, Switzerland (2015).

\bibitem{NOO-rev}
S. Nojiri, S. D. Odintsov and V. K. Oikonomou, {\em Modified Gravity Theories on a Nutshell: 
Inflation, Bounce and Late-time Evolution}, Phys. Rept. {\bf 692} (2017) 1, [1705.11098].


\bibitem{fuj-st} 
Y. Fujii and K. Maeda, {\em The Scalar-Tensor Theory of Gravitation}, Cambridge Monographs on 
Mathematical Physics, Cambridge University Press, United Kingdom (2003).

\bibitem{frni-st}
V. Faraoni, {\em Cosmology in Scalar-Tensor Gravity}, Kluwer Academic, Dordrecht (2004).

\bibitem{ENO-st}
E. Elizalde, S. Nojiri and S. D. Odintsov, {\em Late-time cosmology in (phantom) scalar-tensor 
theory: Dark energy and the cosmic speed-up}, Phys. Rev. {\bf D 70} (2004) 043538, [hep-th/0405034].

\bibitem{FTBM-st}
T. Faulkner, M. Tegmark, E. F. Bunn and Y. Mao, {\em Constraining $f(R)$ Gravity as a Scalar-Tensor 
Theory}, Phys. Rev. {\bf D 76} (2007) 063505, [astro-ph/0612569].

\bibitem{BGP-st}
B. Boisseau, H. Giacomini and D. Polarski, {\em Bouncing Universes in Scalar-Tensor Gravity around
Conformal Invariance}, JCAP {\bf 1605} (2016) 048, [1603.06648].


\bibitem{CM-mm}
A. H. Chamseddine and V. Mukhanov, {\em Mimetic Dark Matter}, JHEP {\bf 1311} (2013) 135, [1308.5410].

\bibitem{CMV-mm}
A. H. Chamseddine, V. Mukhanov and A. Vikman, {\em Cosmology with Mimetic Matter}, JCAP {\bf 1406} 
(2014) 017, [1403.3961].

\bibitem{MV-mm}
L. Mirzagholi and A. Vikman, {\em Imperfect Dark Matter}, JCAP {\bf 1506} (2015) 028, [1412.7136].

\bibitem{HV-mm}
K. Hammer and A. Vikman, {\em Many Faces of Mimetic Gravity}, 1512.09118.

\bibitem{barv-mm}
A. O. Barvinsky, {\em Dark matter as a ghost free conformal extension of Einstein's theory}, JCAP
{\bf 1401} (2014) 014, [1311.3111].

\bibitem{CR-mm}
F. Capela and S. Ramazanov, {\em Modified Dust and the Small Scale Crisis in CDM}, JCAP {\bf 1504}
(2015) 051, [1412.2051].

\bibitem{ramz-mm}
S. Ramazanov, {\em Initial Conditions for Imperfect Dark Matter}, JCAP {\bf 1512} (2015) 007,
[1507.00291].

\bibitem{BR-mm}
E. Babichev and S. Ramazanov, {\em Gravitational focusing of Imperfect Dark Matter}, Phys. Rev. 
{\bf D 95} (2017) no.2, 024025, [1609.08580]. 

\bibitem{SVM-mm}
L. Sebastiani, S. Vagnozzi and R. Myrzakulov, {\em Mimetic gravity: a review of recent
developments and applications to cosmology and astrophysics}, Adv. High Energy Phys. {\bf 2017}
(2017) 3156915, [1612.08661].


\bibitem{bek-dt}
J. D. Bekenstein, {\em The Relation between physical and gravitational geometry}, Phys. Rev. 
{\bf D 48} (1993) 3641, [gr-qc/9211017].

\bibitem{FG-dt}
F. T. Falciano and E. Goulart, {\em A new symmetry of the relativistic wave equation}, Class. 
Quant. Grav. {\bf 29} (2012) 085011, [1112.1341].

\bibitem{BT-dt}
D. Bettoni and S. Liberati, {\em Disformal invariance of second order scalar-tensor theories: 
framing the Horndeski action}, Phys. Rev. {\bf D 88} (2013) 084020, [1306.6724].

\bibitem{DR-dt}
N. Deruelle and J. Rua, {\em Disformal Transformations, Veiled general relativity and Mimetic
Gravity}, JCAP {\bf 1409} (2014) 002, [1407.0825].



\bibitem{RMMM-mdyn}
M. Raza, K. Myrzakulov, D. Momeni and R. Myrzakulov, {\em Mimetic Attractors}, Int. J. Theor.
Phys. {\bf 55} (2016) 2558, [1508.00971].

\bibitem{DKSTS-mdyn}
J. Dutta, W. Khyllep, E. N. Saridakis, N. Tamanini and S. Vagnozzi, {\em Cosmological dynamics 
of mimetic gravity}, JCAP {\bf 1802} (2018) 041, [1711.07290].

\bibitem{LS-mdyn}
G. Leon and E. N. Saridakis, {\em Dynamical behavior in mimetic $F(R)$ gravity}, JCAP {\bf 1504}
(2015) 031, [1501.00488].

\bibitem{OO-mdyn}
S. D. Odintsov and V. K. Oikonomou, {\em Accelerating Cosmology and Phase Structure of $F(R)$ 
Gravity with Lagrange Multiplier Constraint: Mimetic Approach}, Phys. Rev. {\bf D 93} (2016) 
no.2, 023517, [1511.04559].


\bibitem{NO-mfR}
S. Nojiri and S. D. Odintsov, {\em Mimetic $F(R)$ gravity: inflation, dark energy and bounce}, Mod.
Phys. Lett. {\bf A 29} (2014) 1450211, [1408.3561].

\bibitem{NOO-mfR}
S. Nojiri, S. D. Odintsov and V. K. Oikonomou, {\em Viable Mimetic Completion of Unified Inflation-Dark 
Energy Evolution in Modified Gravity}, Phys. Rev. {\bf D 94} (2016) no.10, 104050, [1608.07806].

\bibitem{MSV-mfR}
R. Myrzakulov, L. Sebastiani and S. Vagnozzi, {\em Inflation in $f (R, \f)$ - theories and mimetic
gravity scenario}, Eur. Phys. J. {\bf C 75} (2015) 444, [1504.07984].

\bibitem{OO-mfR1}
S. D. Odintsov and V. K. Oikonomou, {\em Unimodular Mimetic $F(R)$ Inflation}, Astrophys. Space
Sci. {\bf 361} (2016) 236, [1602.05645].

\bibitem{OO-mfR2}
S. D. Odintsov and V. K. Oikonomou, {\em The reconstruction of $f(\vph) R$ and mimetic gravity from 
viable slow-roll inflation}, Nucl. Phys. {\bf B 929} (2018) 79, [1801.10529].


\bibitem{CM-msing}
A. H. Chamseddine and V. Mukhanov, Resolving Cosmological Singularities, JCAP 1703 (2017) 009, 
[1612.05860].

\bibitem{BGY-msing}
S. Brahma, A. Golovnev and D.-H. Yeom, {\em On singularity-resolution in mimetic gravity}, Phys. 
Lett. {\bf B 782} (2018) 280, [1803.03955].

\bibitem{HSP-msing}
J. De Haro, L. A. Sal\'o and Supriya Pan, {\em Limiting curvature mimetic gravity and its relation 
to Loop Quantum Cosmology }, Gen. Rel. Grav. {\bf 51} (2019) no. 4, 49, [1803.09653].


\bibitem{MSVZ-mgal}
R. Myrzakulov, L. Sebastiani, S. Vagnozzi and S. Zerbini, {\em Static spherically symmetric
solutions in mimetic gravity: rotation curves and wormholes}, Class. Quant. Grav. {\bf 33} (2016)
125005, [1510.02284].

\bibitem{vag-mgal}
S. Vagnozzi, {\em Recovering a MOND-like acceleration law in mimetic gravity}, Class. Quant.
Grav. {\bf 34} (2017) 185006, [1708.00603].


\bibitem{MS-mco}
R. Myrzakulov and L. Sebastiani, {\em Spherically symmetric static vacuum solutions in Mimetic
gravity}, Gen. Rel. Grav. {\bf 47} (2015) 89, [1503.04293].

\bibitem{MMGM-mco}
D. Momeni, P. H. R. S. Moraes, H. Gholizade and R. Myrzakulov, {\em Mimetic Compact Stars},
Int. J. Geom. Meth. Mod. Phys. {\bf 15} (2018) no.06, 1850091, [1505.05113].

\bibitem{AO-mco}
A. V. Astashenok and S. D. Odintsov, {\em From neutron stars to quark stars in mimetic gravity},
Phys. Rev. {\bf D 94} (2016) 063008, [1512.07279].

\bibitem{CM-mco}
A. H. Chamseddine and V. Mukhanov, {\em Nonsingular Black Hole}, Eur. Phys. J. {\bf C 77} (2017)
183, [1612.05861].

\bibitem{nas-mco}
G. G. L. Nashed, {\em Spherically symmetric black hole solution in mimetic gravity and anti-evaporation},
Int. J. Geom. Meth. Mod. Phys. {\bf 15} (2018) no.09, 1850154.

\bibitem{NHB-mco}	
G. G. L. Nashed, W. El Hanafy and K. Bamba, {\em Charged rotating black holes coupled with nonlinear 
electrodynamics Maxwell field in the mimetic gravity}, JCAP {\bf 1901} (2019) no.01, 058, [1809.02289].


\bibitem{SJ-mgw}
J. Sakstein and B. Jain, {\em Implications of the Neutron Star Merger GW170817 for Cosmological
Scalar-Tensor Theories}, Phys. Rev. Lett. {\bf 119} (2017) no.25, 251303, [1710.05893].

\bibitem{BBFLMS-mgw}
T. Baker, E. Bellini, P. G. Ferreira, M. Lagos, J. Noller and I. Sawicki, {\em Strong constraints 
on cosmological gravity from GW170817 and GRB 170817A}, Phys. Rev. Lett. {\bf 119} (2017) no.25,
251301, [1710.06394].

\bibitem{LSYN-mgw}
D. Langlois, R. Saito, D. Yamauchi and K. Noui, {\em Scalar-tensor theories and modified gravity
in the wake of GW170817}, Phys. Rev. {\bf D 97} (2018) no.6, 061501, [1711.07403].

\bibitem{BPT-mgw}
R. A. Battye, F. Pace and D. Trinh, {\em Gravitational wave constraints on dark sector models},
Phys. Rev. {\bf D 98} (2018) no.2, 023504, [1802.09447].

\bibitem{RSCV-mgw}
M. Rinaldi, L. Sebastiani, A. Casalino and S. Vagnozzi, {\em Mimicking dark matter and dark
energy in a mimetic model compatible with GW170817},  Phys. Dark Univ. {\bf 22} (2018) 108, 
[1803.02620].

\bibitem{GBKM-mgw}
A. Ganz, N. Bartolo, P. Karmakar and S. Matarrese, {\em Gravity in mimetic scalar-tensor theories 
after GW170817}, JCAP {\bf 1901} (2019) no.01, 056, [1809.03496].

\bibitem{CRSV-mgw}
A. Casalino, M. Rinaldi, L. Sebastiani and S. Vagnozzi, {\em Alive and well: mimetic gravity and a 
higher-order extension in light of GW170817}, Class. Quant. Grav. {\bf 36} (2019) no.1, 017001,
[1811.06830].


\bibitem{LSV-dde}
E. A. Lim, I. Sawicki and A. Vikman, {\em Dust of dark energy}, JCAP {\bf 1005} (2010) 012, 
[1003.5751].

\bibitem{GGWC-dde}
C. Gao, Y. Gong, X. Wang and X. Chen, {\em Cosmological models with Lagrange multiplier field}, 
Phys. Lett. {\bf B 702} (2011) no.2-3, 107, [1003.6056].

\bibitem{CMNO-dde}
S. Capozziello, J. Matsumoto, S. Nojiri and S. D. Odintsov, {\em Dark energy from modified gravity 
with Lagrange multipliers}, Phys. Lett. {\bf B 693} (2010) no.2, 198, [1004.3691].


\bibitem{CP-mhd}
Y. Cai and Y.-S. Piao, {\em Higher order derivative coupling to gravity and its cosmological
implications}, Phys. Rev. {\bf D 96} (2017) no.12, 124028, [1707.01017].

\bibitem{GMF-mhd}
M. A. Gorji, S. A. H. Mansoori and H. Firouzjahi, {\em Higher Derivative Mimetic Gravity},
JCAP {\bf 1801} (2018) 020 [1709.09988].


\bibitem{CKOT-minst}
M. Chaichian, J. Kluson, M. Oksanen and A. Tureanu, {\em Mimetic dark matter, ghost instability
and a mimetic tensor-vector-scalar gravity}, JHEP {\bf 1412} (2014) 102, [1404.4008].

\bibitem{ZSML-minst}
Y. Zheng, L. Shen, Y. Mou and M. Li, {\em On (in)stabilities of perturbations in mimetic models
with higher derivatives}, JCAP {\bf 1708} (2017) 040, [1704.06834].

\bibitem{HNK-minst}
S. Hirano, S. Nishi and T. Kobayashi, {\em Healthy imperfect dark matter from effective theory of
mimetic cosmological perturbations}, JCAP {\bf 1707} (2017) 009, [1704.06031].

\bibitem{TK-minst}
K. Takahashi and T. Kobayashi, {\em Extended mimetic gravity: Hamiltonian analysis and gradient
instabilities}, JCAP {\bf 1711} (2017) 038, [1708.02951].


\bibitem{BLN-mdhost}
J. Ben Achour, D. Langlois and K. Noui, {\em Degenerate higher order scalar-tensor theories beyond 
Horndeski and disformal transformations}, Phys. Rev. {\bf D 93} (2016) 124005, [1602.08398].

\bibitem{LMNV-mdhost}
D. Langlois, M. Mancarella, K. Noui, F. Vernizzi, {\em Mimetic gravity as DHOST theories}, 1802.03394.


\bibitem{MS-mnfR}
R. Myrzakulov and L. Sebastiani, {\em Non-local $F(R)$ - mimetic gravity}, Astrophys. Space Sci. 
{\bf 361} (2016) 188, [1601.04994].

\bibitem{MMG-mfRv}
D. Momeni, R. Myrzakulov and E. Gudekli, {\em Cosmological viable mimetic $f (R)$ and $f (R, T)$
theories via Noether symmetry}, Int. J. Geom. Meth. Mod. Phys. {\bf 12} (2015) 1550101, [1502.00977].

\bibitem{AOO-mfRv}
A. V. Astashenok, S. D. Odintsov and V. K. Oikonomou, {\em Modified Gauss-Bonnet gravity with
the Lagrange multiplier constraint as mimetic theory}, Class. Quant. Grav. {\bf 32} (2015) 185007,
[1504.04861].

\bibitem{BHHA-mfRv}
E. H. Baffou, M. J. S. Houndjo, M. Hamani-Daouda and F. G. Alvarenga, {\em Late time cosmological 
approach in mimetic $f(R,T)$ gravity}, Eur. Phys. J. {\bf C 77} (2017) no.10, 708, [1706.08842].


\bibitem{ABKM-mh}
F. Arroja, N. Bartolo, P. Karmakar and S. Matarrese, {\em The two faces of mimetic Horndeski gravity: 
disformal transformations and Lagrange multiplier}, JCAP {\em 1509} (2015) 051, [1506.08575].

\bibitem{CMSVZ-mh}
G. Cognola, R. Myrzakulov, L. Sebastiani, S. Vagnozzi and S. Zerbini, {\em Covariant Horava-like and 
mimetic Horndeski gravity: cosmological solutions and perturbations}, Class. Quant. Grav. {\bf 33} 
(2016) 225014, [1601.00102].


\bibitem{BCC-mbi}
M. Bouhmadi-Lopez, C.-Y. Chen and P. Chen, {\em Primordial Cosmology in Mimetic Born-Infeld Gravity}, 
JCAP {\bf 1711} (2017) 053, [1709.09192].

\bibitem{CBC-mbi}
C.-Y. Chen, M. Bouhmadi-Lopez and P. Chen, {\em Black hole solutions in mimetic Born-Infeld gravity},  
Eur. Phys. J. {\bf C 78} (2018) no.1, 59, [1710.10638].


\bibitem{SN-mbw}
N. Sadeghnezhad and K. Nozari, {\em Braneworld Mimetic Cosmology}, Phys. Lett. {\bf B 769} (2017) 134, 
[1703.06269].

\bibitem{YYZL-mbw}
Y. Zhong, Y. Zhong, Y.-P. Zhang, Y.-X. Liu, {\em Thick branes with inner structure in mimetic gravity}, 
Eur. Phys. J. {\bf C 78} (2018) 45, [1711.09413].

\bibitem{ZZGL-mbw}
Y. Zhong, Y.-P. Zhang, W.-D. Guo and Y.-X. Liu, {\em Gravitational resonances in mimetic thick branes},
JHEP {\bf 1904} (2019) 154, [1812.06453].


\bibitem{CM-mmg1}
A. H. Chemseddine and V. Mukhanov, {\em Ghost Free Mimetic Massive Gravity}, JHEP  {\bf 1806} (2018)
060, [1805.06283].

\bibitem{CM-mmg2}
A. H. Chemseddine and V. Mukhanov, {\em Mimetic Massive Gravity: Beyond Linear Approximation}, JHEP  
{\bf 1806} (2018) 062, [1805.06598].

\bibitem{MS-mmg}
O. Malaeb and C. Saghir, {\em Hamiltonian Formulation of Ghost Free Mimetic Massive Gravity}, Eur. Phys. J. 
{\bf C 79} (2019) no.7, 584, [1901.06727].

\bibitem{SVA-mmg}
A. R. Solomon, V. Vardanyan and Y. Akrami, {\em Massive mimetic cosmology}, Phys. Lett. {\bf B 794} (2019) 
135, [1902.08533]. 


\bibitem{MAM-mfR}
D. Momeni, A. Altaibayeva and R. Myrzakulov, {\em New Modified Mimetic Gravity}, Int. J. Geom. Meth. 
Mod. Phys. {\bf 11} (2014) 1450091, [1407.5662].

\bibitem{MSVZ-mhor}
R. Myrzakulov, L. Sebastiani, S. Vagnozzi and S. Zerbini, {\em Mimetic covariant renormalizable gravity}, 
Fund. J. Mod. Phys. {\bf 8} (2015) 119–124, [1505.03115].

\bibitem{kosh-mhd}
N. A. Koshelev, {\em Effective dark matter fluid with higher derivative corrections}, 1512.07097.

\bibitem{ST-m2s}
E. N. Saridakis and M. Tsoukalas, {\em Bi-scalar modified gravity and cosmology with conformal invariance}, 
JCAP {\bf 1604} (2016) 017, [1602.06890].

\bibitem{KNY-mvt}
R. Kimura, A. Naruko and D. Yoshida, {\em Extended vector-tensor theories}, JCAP {\bf 1701} (2017) 002, 
[1608.07066].

\bibitem{GMFM-mgf}
M. A. Gorji, S. Mukohyama, H. Firouzjahi and S. A. Hosseini Mansoori, {\em Gauge Field Mimetic Cosmology},
JCAP {\bf 1808} (2018) no.08, 047, [1807.06335].

\bibitem{JV-mgf}
P. Jirou\v{s}ek and A. Vikman, {\em New Weyl-invariant vector-tensor theory for the cosmological constant}, 
JCAP {\bf 1904} (2019) 004, [1811.09547].


\bibitem{HHSS-mmEA-1}
Z. Haghani, T. Harko, H. R. Sepangi and S. Shahidi, {\em The scalar Einstein-Aether theory}, 1404.7869.

\bibitem{JS-mmEA-1}
T. Jacobson and A. J. Speranza, {\em Comment on ``Scalar Einstein-Aether theory"}, 1405.6351.

\bibitem{HHSS-mmEA-2}
Z. Haghani, T. Harko, H. R. Sepangi and S. Shahidi, {\em Cosmology of a Lorentz violating Galileon theory},
JCAP {\bf 1505} (2015) 022, [1501.00819].

\bibitem{JS-mmEA-2}
T. Jacobson and A. J. Speranza, {\em Variations on an aethereal theme}, Phys. Rev. {\bf D 92} (2015) 044030,
[1503.08911]. 

\bibitem{sper-mmEA}
A. J. Speranza, {\em Ponderable aether}, JCAP {\bf 1508} (2015) 016, [1504.03305].

\bibitem{RACMP-mhorlif}
S. Ramazanov, F. Arroja, M. Celoria, S. Matarrese and L. Pilo, {\em Living with ghosts in Horava-Lifshitz 
gravity}, JHEP {\bf 06} (2016) 020, [1601.05405].


\bibitem{traut}
A. Trautman, {\em Spin and torsion may avert gravitational singularities}, Nature {\bf 242} (1973) 7.

\bibitem{HVKN-trev}
F. W. Hehl, P. Von Der Heyde, G. Kerlick and J. Nester, {\em General Relativity with Spin
and Torsion: Foundations and Prospects}, Rev. Mod. Phys. {\bf 48} (1976) 393.

\bibitem{akr-tbook}
A. K. Raychaudhuri, {\em Theoretical Cosmology}, Clarendon Press, Oxford, United Kingdom (1979).

\bibitem{SG-tbook}
V. de Sabbata and M. Gasperini, {\em Introduction to Gravitation}, World Scientific, Singapore (1985).

\bibitem{SS-tbook}
V. de Sabbata and C. Sivaram, {\em Spin Torsion and Gravitation}, World Scientific, Singapore (1994).

\bibitem{HDMN-trev}
F. W. Hehl, J. D. McCrea, E. W. Mielke and Y. Ne\'eman, {\em Metric affine gauge theory of gravity:
Field equations, Noether identities, world spinors and breaking of dilation invariance}, Phys.
Rept. {\bf 258} (1995) 1, [gr-qc/9402012].

\bibitem{HO-trev}
F. W. Hehl and Y. N. Obukhov, {\em How does the electromagnetic field couple to gravity, in
particular to metric, nonmetricity, torsion and curvature?}, Lect. Notes Phys. {\bf 562} (2001) 479,
[gr-qc/0001010].

\bibitem{shap-trev}
I. L. Shapiro, {\em Physical aspects of the space-time torsion}, Phys. Rept. {\bf 357} (2002) 113,
[hep-th/0103093].

\bibitem{blag-book}
M. Blagojevic, {\em Gravitation and Gauge symmetries}, IOP Publishing, London, United Kingdom (2002).

\bibitem{fab-thesis}
L. Fabbri, {\em Higher-Order Theories of Gravitation}, Ph.D. Thesis, 0806.2610.

\bibitem{CL-extgrav}
S. Capozziello and M. De Laurentis, {\em Extended Theories of Gravity}, Phys. Rept. {\bf 509} (2011)
167, [1108.6266].

\bibitem{popl-trev}
N. Poplawski, {\em Affine theory of gravitation}, Gen. Rel. Grav. {\bf 46} (2014) 1625, [1203.0294].

\bibitem{WZ-trev}
H. F. Westman and T. G. Zlosnik, {\em An introduction to the physics of Cartan gravity},
Annals Phys. {\bf 361} (2015) 330, [1411.1679].


\bibitem{pmssg}
P. Majumdar and S. SenGupta, {\em Parity violating gravitational coupling of electromagnetic fields},
Class. Quant. Grav. {\bf 16} (1999) L89, [gr-qc/9906027].

\bibitem{ham}
R. T. Hammond, {\em Strings in gravity with torsion}, Gen. Rel. Grav. 32 (2000) 2007, [gr-qc/9904033].


\bibitem{ssgss-kr1}
S. SenGupta and S. Sur, {\em Spherically symmetric solutions of gravitational field equations in
Kalb-Ramond background}, Phys. Lett. {\bf B 521} (2001) 350, [gr-qc/0102095].

\bibitem{ssgas-kr}
S. SenGupta and A. Sinha, {\em Fermion helicity flip in a Kalb-Ramond background}, Phys. Lett. 
{\bf B 514} (2001) 109, [hep-th/0102073].

\bibitem{DJM-kr}
P. Das, P. Jain and S. Mukherji, {\em Cosmic birefringence within the framework of heterotic string
theory}, Int. J. Mod. Phys. {\bf A 16} (2001) 4011, [hep-ph/0011279].

\bibitem{skpmssgas-kr}
S. Kar, P. Majumdar, S. SenGupta and A. Sinha, {\em Does a Kalb-Ramond field make space-time
optically active?}, Eur. Phys. J. {\bf C 23} (2002) 357, [gr-qc/0006097].

\bibitem{skpmssgss-kr}
S. Kar, P. Majumdar, S. SenGupta and S. Sur, {\em Cosmic optical activity from an inhomogeneous
Kalb-Ramond field}, Class. Quant. Grav. {\bf 19} (2002) 677, [hep-th/0109135].

\bibitem{skssgss-kr}
S. Kar, S. SenGupta and S. Sur, {\em Static spherisymmetric solutions, gravitational lensing and
perihelion precession in Einstein-Kalb-Ramond theory}, Phys. Rev. {\bf D 67} (2003) 044005,
[hep-th/0210176].

\bibitem{ssgss-kr2}
S. SenGupta and S. Sur, {\em Does curvature dilaton coupling with a Kalb-Ramond field lead to an
accelerating universe?}, JCAP {\bf 0312} (2003) 001, [hep-th/0207065].

\bibitem{dmssgss-kr1}
D. Maity, S. SenGupta and S. Sur, {\em Spinning test particle in Kalb-Ramond background}, Eur.
Phys. J. {\bf C 42} (2005) 453, [hep-th/0409143].

\bibitem{sssdssg-kr}
S. Sur, S. Das and S. SenGupta, {\em Charged black holes in generalized dilaton-axion gravity},
JHEP {\bf 0510} (2005) 064, [hep-th/0508150].

\bibitem{AMT-kr}
J. Alexandre, N. E. Mavromatos and D. Tanner, {\em Non-perturbative time-dependent String Backgrounds 
and Axion-induced Optical Activity}, Phys. Rev. {\bf D 78} (2008) 066001, [arXiv:0804.2353].

\bibitem{BC-kr}
S. Bhattacharjee and A. Chatterjee, {\em Gauge invariant coupling of fields to torsion: a string
inspired model}, Phys. Rev. {\bf D 83} (2011) 106007, [1101.0118].

\bibitem{CMS-kr}
M. de Cesare, N. E. Mavromatos and S. Sarkar, On the possibility of tree-level leptogenesis from 
Kalb-Ramond torsion background, Eur. Phys. J. C 75 (2015) 514, [arXiv:1412.7077].


\bibitem{bmssssg-kr1}
B. Mukhopadhyaya, S. Sen and S. SenGupta, {\em Does a Randall-Sundrum scenario create the
illusion of a torsion free universe?}, Phys. Rev. Lett. {\bf 89} (2002) 121101 [{\it Erratum 
ibid.} {\bf 89} (2002) 259902], [hep-th/0204242].

\bibitem{ssgss-kr3}
S. SenGupta and S. Sur, {\em Gravitational Redshift in Einstein-Kalb-Ramond space-time and
Randall-Sundrum scenario}, Europhys. Lett. {\bf 65} (2004) 601, [hep-th/0306048].

\bibitem{dmssg-kr}
D. Maity and S. SenGupta, {\em Cosmic optical activity in a Randall-Sundrum brane world with
bulk Kalb-Ramond field}, Class. Quant. Grav. {\bf 21} (2004) 3379 [hep-th/0311142].

\bibitem{dmssgss-kr2}
D. Maity, S. SenGupta and S. Sur, {\em Observable signals in a string inspired axion-dilaton
background and Randall-Sundrum scenario}, Phys. Rev. {\bf D 72} (2005) 066012, [hep-th/0507210].

\bibitem{sdadssg-kr}
S. Das, A. Dey and S. SenGupta, {\em Readdressing the hierarchy problem in a Randall-Sundrum scenario 
with bulk Kalb-Ramond background}, Class. Quant. Grav. {\bf 23} (2006) L67, [hep-th/0511247].

\bibitem{bmssssg-kr2}
B. Mukhopadhyaya, S. Sen and S. SenGupta, {\em A Randall-Sundrum scenario with bulk dilaton and torsion}, 
Phys. Rev. {\bf D 79} (2009) 124029, [0903.0722].

\bibitem{adbmssg-kr}
A. Das, B. Mukhopadhyaya and S. SenGupta, {\em Why has spacetime torsion such negligible effect on the 
Universe?}, Phys. Rev. {\bf D 90} (2014) 107901, [1410.0814].

\bibitem{scssg-kr1}
S. Chakraborty and S. SenGupta, {\em Solutions on a brane in a bulk spacetime with Kalb-Ramond field}, 
Annals Phys. {\bf 367} (2016) 258, [1412.7783].

\bibitem{scssg-kr2}
S. Chakraborty and S. SenGupta, {\em Strong gravitational lensing -- A probe for extra dimensions and 
Kalb-Ramond field}, JCAP {\bf 1707} (2017) no.07, 045, [1611.06936].

\bibitem{scssg-kr3}
S. Chakraborty and S. SenGupta, {\em Packing extra mass in compact stellar structures: An interplay 
between Kalb-Ramond field and extra dimensions}, JCAP {\bf 1805} (2018) no.05, 032, [1708.08315].


\bibitem{HRR-proptor}
S. Hojman, M. Rosenbaum and M.P. Ryan, {\em Propagating torsion and gravitation}, Phys. Rev. {\bf D 19} 
(1979) 430.

\bibitem{CF-proptor}
S.M. Carroll and G.B. Field, {\em Consequences of propagating torsion in connection dynamic theories of 
gravity}, Phys. Rev. {\bf D 50} (1994) 3867, [gr-qc/9403058].

\bibitem{saa-proptor}
A. Saa, {\em Propagating torsion from first principles}, Gen. Rel. Grav. {\bf 29} (1997) 205,
[gr-qc/9609011].

\bibitem{BS-proptor}
A.S. Belyaev and I.L. Shapiro, {\em The action for the (propagating) torsion and the limits on the torsion 
parameters from present experimental data}, Phys. Lett. {\bf B 425} (1998) 246, [hep-ph/9712503].

\bibitem{popl-proptor}
N.J. Poplawski, {\em Propagating torsion in the Einstein gauge}, J. Math. Phys. {\bf 47} (2006) 112504, 
[gr-qc/0605061].

\bibitem{BC-proptor}
M. Blagojevic and B. Cvetkovic, {\em Three-dimensional gravity with propagating torsion: Hamiltonian 
structure of the scalar sector}, Phys. Rev. {\bf D 88} (2013) 104032, [1309.0411].

\bibitem{ND-proptor}
V. Nikiforova and T. Damour, {\em Infrared modified gravity with propagating torsion: instability of 
torsionfull de Sitter-like solutions}, Phys. Rev. {\bf D 97} (2018) no.12, 124014, [1804.09215].


\bibitem{HMS-pvtor}
R. Hojman, C. Mukku and W.A. Sayed, {\em Parity violation in metric torsion theories of gravitation}, 
Phys. Rev. {\bf D 22} (1980) 1915.

\bibitem{holst-pvtor}
S. Holst, {\em Barbero's Hamiltonian derived from a generalized Hilbert-Palatini action}, Phys. Rev. 
{\bf D 53} (1996) 5966, [gr-qc/9511026].

\bibitem{bmssgss-pvtor}
B. Mukhopadhyaya, S. SenGupta and S. Sur, {\em Space-time torsion and parity violation: A gauge invariant 
formulation}, Mod. Phys. Lett. {\bf A 17} (2002) 43, [hep-th/0106236].

\bibitem{bmssssgss-pvtor}
B. Mukhopadhyaya, S. Sen, S. SenGupta and S. Sur, {\em Parity violation and torsion: A study in 
four-dimensions and higher dimensions}, Eur. Phys. J. {\bf C 35} (2004) 129, [hep-th/0207165].

\bibitem{dmpmssg-pvtor}
D. Maity, P. Majumdar and S. SenGupta, {\em Parity violating Kalb-Ramond-Maxwell interactions and CMB 
anisotropy in a brane world}, JCAP {\bf 0406} (2004) 005, [hep-th/0401218].

\bibitem{CM-pvtor}
A. Chatterjee and P. Majumdar, {\em Kalb-Ramond field interactions in a braneworld scenario}, Phys. Rev. 
{\bf D 72} (2005) 066013, [hep-th/0507085].

\bibitem{FMT-pvtor}
L. Freidel, D. Minic and T. Takeuchi, {\em Quantum gravity, torsion, parity violation and all that}, 
Phys. Rev. {\bf D 72} (2005) 104002, [hep-th/0507253].


\bibitem{ACCF-extgr}
G. Allemandi, M. Capone, S. Capozziello and M. Francaviglia, {\em Conformal aspects of Palatini
approach in extended theories of gravity}, Gen. Rel. Grav. {\bf 38} (2006) 33, [hep-th/0409198].

\bibitem{FC-extgr}
V. Faraoni and S. Capozziello, {\em Beyond Einstein Gravity: A Survey of Gravitational Theories for 
Cosmology and Astrophysics}, Fundamental Theories of Physics {\bf 170} (2010), Dordrecht, Springer (2011).

\bibitem{CMT-extgr}
D.A. Carranza, S. Mendoza and L.A. Torres, {\em A cosmological dust model with extended $f(\chi)$ gravity}, 
Eur. Phys. J. {\bf C 73} (2013) 2282, [1208.2502].


\bibitem{ROH-skew}
G.F. Rubilar, Y.N. Obukhov and F.W. Hehl, {\em Torsion nonminimally coupled to the electromagnetic field 
and birefringence}, Class. Quant. Grav. {\bf 20} (2003) L185, [gr-qc/0305049].

\bibitem{HORB-skew}
F.W. Hehl, Yu. N. Obukhov, G.F. Rubilar and M. Blagojevic, {\em On the theory of the skewon field: From 
electrodynamics to gravity}, Phys. Lett. {\bf A 347} (2005) 14, [gr-qc/0506042].

\bibitem{Ni-skew}
W.-T. Ni, {\em Skewon field and cosmic wave propagation}, Phys. Lett. {\bf A 378} (2014) 1217, [1312.3056].


\bibitem{FRM-scaltor}
J. B. Fonseca-Neto, C. Romero and S. P. G. Martinez, {\em Scalar torsion and a new symmetry of general 
relativity}, Gen. Rel. Grav. {\bf 45} (2013) 1579, [1211.1557].


\bibitem{VFS-sqtor}
S. Vignolo, L. Fabbri and C. Stornaiolo, {\em A square-torsion modification of Einstein-Cartan theory}, 
Annalen Phys. {\bf 524} (2012) 826, [1201.0286].

\bibitem{lu-sqtor}
J.-A. Lu, {\em $R + S^2$ theories of gravity without big-bang singularity}, Annals Phys. {\bf 354} (2015) 424,
[1404.4440].

\bibitem{fab-sqtor}
L. Fabbri, {\em A discussion on the most general torsion-gravity with electrodynamics for Dirac spinor matter 
fields}, Int. J. Geom. Meth. Mod. Phys. {\bf 12} (2015) no. 09, 1550099, [1409.2007].

\bibitem{VCVM-sqtor}
T. B. Vasilev, J. A. R. Cembranos, J. G. Valcarcel and P. Martín-Moruno, {\em Stability in quadratic torsion 
theories}, Eur. Phys. J. {\bf C 77} (2017) no. 11, 755, [1706.07080].   


\bibitem{KS-deg1}
R. K. Kaul and S. Sengupta, {\em Degenerate spacetimes in first order gravity}, Phys. Rev. {\bf D 93}
(2016) no.8, 084026, [1602.04559].

\bibitem{KS-deg2}
R. K. Kaul and S. Sengupta, {\em New solutions in pure gravity with degenerate tetrads}, Phys. Rev. 
{\bf D 94} (2016) no.10, 104047, [1609.02344].

\bibitem{SS-deg1}
S. Sengupta, {\em Spacetime-bridge solutions in vacuum gravity}, Phys. Rev. {\bf D 96} (2017) no.10, 
104031, [1708.04971].

\bibitem{KS-deg3}
R. K. Kaul and S. Sengupta, {\em Degenerate extension of the Schwarzschild exterior}, Phys. Rev. 
{\bf D 96}  (2017) no.10, 104011, [1709.00188].

\bibitem{SS-deg2}
S. Sengupta, {\em Time travel in vacuum spacetimes}, Phys. Rev. {\bf D 97} (2018) no.12, 124038,
[1805.03035].


\bibitem{ssasb-mst1}
S. Sur and A. S. Bhatia, {\em Weakly dynamic dark energy via metric-scalar couplings with torsion},
JCAP {\bf 1707} (2017) 039, [1611.00654].

\bibitem{ssasb-mst2}
A. S. Bhatia and S. Sur, {\em Phase Plane Analysis of Metric-Scalar Torsion Model for Interacting 
Dark Energy}, 1611.06902.

\bibitem{ssasb-mst3}
A. S. Bhatia and S. Sur, {\em Dynamical system analysis of dark energy models in scalar coupled 
metric-torsion theories}, Int. J. Mod. Phys. {\bf D 26} (2017) no.13, 1750149, [1702.01267].


\bibitem{FF-fT}
R. Ferraro and F. Fiorini, {\em Modified teleparallel gravity: Inflation without inflaton}, Phys. 
Rev. {\bf D 75} (2007) 084031, [gr-qc/0610067].

\bibitem{BF-fT}
G. R. Bengochea and R. Ferraro, {\em Dark torsion as the cosmic speed-up}, Phys. Rev. {\bf D 79},
(2009) 124019, [0812.1205]. 

\bibitem{LSB-fT}
B. Li, T.P. Sotiriou and J. D. Barrow, {\em Large-scale Structure in $f(T)$ Gravity}, Phys. Rev. 
{\bf D 83} (2011) 104017, [1103.2786].

\bibitem{CCDDS-fT}
Y.-F. Cai, S.-H. Chen, J. B. Dent, S. Dutta and E. N. Saridakis, {\em Matter Bounce Cosmology with 
the $f(T)$ Gravity}, Class. Quant. Grav. {\bf 28} (2011) 215011, [1104.4349].

\bibitem{BMT-fT}
C. G. B\"{o}hmer, A. Mussa and N. Tamanini, {\em Existence of relativistic stars in $f(T)$ gravity},
Class. Quant. Grav. {\bf 28} (2011) 245020, [1107.4455].

\bibitem{GLSW-fT}
C.-Q. Geng, C.-C. Lee, E. N. Saridakis and Y.-P. Wu, {\em ``Teleparallel'' dark energy}, Phys. Lett.
{\bf B 704} (2011) 384, [1109.1092].

\bibitem{GLS-fT}
C.-Q. Geng, C.-C. Lee and E. N. Saridakis, {\em Observational Constraints on Teleparallel Dark
Energy}, JCAP {\bf 1201} (2012) 002, [1110.0913].

\bibitem{IS-fT}
L. Iorio and E. N. Saridakis, {\em Solar system constraints on $f(T)$ gravity}, Mon. Not. Roy. 
Astron. Soc. {\bf 427} (2012) 1555, [1203.5781].

\bibitem{CGSV-fT}
S. Capozziello, P. A. Gonzalez, E. N. Saridakis and Y. Vasquez, {\em Exact charged black-hole solutions 
in $D$-dimensional $f(T)$ gravity: torsion vs curvature analysis}, JHEP {\bf 1302} (2013) 039, [1210.1098].

\bibitem{SST-fT}
M. A. Skugoreva, E. N. Saridakis and A. V. Toporensky, {\em Dynamical features of scalar-torsion theories}, 
Phys. Rev. {\bf D 91} (2015) 044023, [1412.1502].

\bibitem{KPS-fT}
G. Kofinas, E. Papantonopoulos and E. N. Saridakis, {\em Self-Gravitating Spherically Symmetric Solutions 
in Scalar-Torsion Theories}, Phys. Rev. {\bf D 91} (2015) 104034, [1501.00365].

\bibitem{CCLS-fT}
Y.-F. Cai, S. Capozziello, M. De Laurentis and E. N. Saridakis, {\em $f(T)$ teleparallel gravity and 
cosmology}, Rept. Prog. Phys. {\bf 79} (2016) 106901, [1511.07586].

\bibitem{BB-fT}
S. Bahamonde and C. G. B\"{o}hmer, {\em Modified teleparallel theories of gravity: Gauss-Bonnet and trace 
extensions}, Eur. Phys. J. {\bf C 76} (2016) no.10, 578, [1606.05557].

\bibitem{BCFN-fT}
S. Bahamonde, S. Capozziello, M. Faizal and R. C. Nunes, {\em Nonlocal Teleparallel Cosmology}, Eur. Phys. 
J. {\bf C 77} (2017) no.9, 628, [1709.02692].

\bibitem{CLPR-fT}
S. Capozziello, O. Luongo, R. Pincak and A. Ravanpak, {\em Cosmic acceleration in non-flat $f(T)$ cosmology}, 
Gen. Rel. Grav. {\bf 50} (2018) no.5, 53, [1804.03649]. 

\bibitem{FSGS-fT}
G. Farrugia, J. Levi Said, V. Gakis and E. N. Saridakis, {\em Gravitational Waves in Modified Teleparallel 
Theories}, Phys. Rev. {\bf D 97} (2018) no.12, 124064, [1804.07365].


\bibitem{YN-PGT}
H.-J. Yo and J. M. Nester, {\em Dynamic Scalar Torsion and an Oscillating Universe}, Mod. Phys. Lett. 
{\bf A 22} (2007) 2057, [astro-ph/0612738].

\bibitem{MGK-PGT}
A. V. Minkevich, A. S. Garkun and V. I. Kudin, {\em Regular accelerating universe without dark energy}, 
Class. Quant. Grav. {\bf 24} (2007) 5835, [0706.1157].

\bibitem{NSV-PGT}
J. M. Nester, L. L. So and T. Vargas, {\em On the energy of homogeneous cosmologies}, Phys. Rev. {\bf D 78} 
(2008) 044035, [0803.0181].

\bibitem{SNY-PGT}
K.-F. Shie, J.M. Nester and H.-J. Yo, {\em Torsion Cosmology and the Accelerating Universe}, Phys. Rev. 
{\bf D 78} (2008) 023522, [0805.3834].

\bibitem{mink-PGT}
A. V. Minkevich, {\em Accelerating Universe without dark energy and dark matter and spacetime torsion}, 
Phys. Lett. {\bf B 678} (2009) 423, [0902.2860].

\bibitem{BHN-PGT}
P. Baekler, F. W. Hehl and J. M. Nester, {\em Poincar\'e gauge theory of gravity: Friedmann cosmology with 
even and odd parity modes. Analytic part}, Phys. Rev. {\bf D 83} (2011) 024001, [1009.5112].

\bibitem{GLT-PGT}
C.-Q. Geng, C.-C. Lee and H.-H. Tseng, {\em Scalar-Torsion Cosmology in the Poincar\'e Gauge Theory of 
Gravity}, JCAP {\bf 1211} (2012) 013, [1207.0579].

\bibitem{HB-PGT}
M. Blagojevic and F. W. Hehl, {\em Gauge Theories of Gravitation: A Reader with Commentaries}, World 
Scientific, Singapore (2013).

\bibitem{LC-PGT}
J. Lu and G. Chee, {\em Cosmology in Poincar\'e gauge gravity with a pseudoscalar torsion}, JHEP {\bf 1605} 
(2016) 024, [1601.03943].

\bibitem{NRR-PGT}
V. Nikiforova, S. Randjbar-Daemi and V. Rubakov, {\em Self-accelerating Universe in modified gravity with 
dynamical torsion}, Phys. Rev. {\bf D 95} (2017) no. 2, 024013, [1606.02565]. 

\bibitem{obu-PGT}
Y. N. Obukhov, {\em Poincar\'e gauge gravity: An overview}, Int. J. Geom. Meth. Mod. Phys. {\bf 15} (2018) 
no. supp.01, 1840005, [1805.07385]. 


\bibitem{KRT-texpt}
V. A. Kostelecky, N. Russell and J. Tasson, {\em New Constraints on Torsion from Lorentz Violation}, 
Phys. Rev. Lett. {\bf 100} (2008) 111102, [0712.4393].

\bibitem{FR-texpt}
E. E. Flanagan and E. Rosenthal, {\em Can gravity probe B usefully constrain torsion gravity theories?}, 
Phys. Rev. {\bf D 75} (2007) 124016, [0704.1447].
 
\bibitem{BF-texpt} 
O. V. Babourova and B. N. Frolov, {\em Interaction of the 4-rotational gauge field with orbital momentum, 
gravidiamagnetic effect and orbit experiment `Gravity Probe B'}, Phys. Rev. {\bf D 82} (2010) 027503, 
[1004.1790].

\bibitem{HOP-texpt}
F. W. Hehl, Y. N. Obukhov and D. Puetzfeld, {\em On Poincar\'e gauge theory of gravity, its equations of 
motion and Gravity Probe B}, Phys. Lett. {\bf A 377} (2013) 1775, [1304.2769].
 
\bibitem{CCR-text} 
S. Camera, V. F. Cardone and N. Radicella, {\em Detectability of Torsion Gravity via Galaxy Clustering and 
Cosmic Shear Measurements}, Phys. Rev. {\bf D 89} (2014) 083520, [1311.1004]. 

\bibitem{CCSZ-texpt}
O. Castillo-Felisola, C. Corral, I. Schmidt and A. R. Zerwekh, {\em Updated limits on extra dimensions 
through torsion and LHC data}, Mod. Phys. Lett. {\bf A 29} (2014) 1450081, [1404.5195].

\bibitem{LP-texpt}
S. Lucat and T. Prokopec, {\em Observing Geometrical Torsion}, arXiv:1705.00889.


\bibitem{MO-mfT}
B. Mirza and F. Oboudiat, {\em Mimetic $f (T)$ Teleparallel gravity and cosmology}, 1712.03363.

\bibitem{GZYSL-mfT}
W.-D. Guo, Y. Zhong, K. Yang, T.-T. Sui and Y.-X. Liu, {\em Thick brane in mimetic $f (T)$ gravity},
1805.05650.


\bibitem{Planck18}
Planck Collaboration: N. Aghanim {\it et. al.}, {\em Planck 2018 results. VI. Cosmological parameters},
1807.06209.


\bibitem{einst-rel}
A. Einstein, {\em The Meaning of Relativity: Fifth edition, including the RELATIVISTIC THEORY OF THE 
NON-SYMMETRIC FIELD}, Princeton University Press, New Jersey, 1970.


\bibitem{mus-MTconf}
J. A. Musante, {\em Conformal Gauge Relativity: On the Geometrical Unification of Gravitation and Gauge 
Fields}, 1008.2677. 

\bibitem{fab-MTconf}
L. Fabbri, {\em Metric-Torsional Conformal Gravity}, Phys. Lett. {\bf B 707} (2012) 415, [1101.1761]. 

\bibitem{berg-MTconf}
D. R. Bergman, {\em Internal Symmetry of Space-Time Connections with Torsion}, 1411.5568.


\bibitem{BOS-TPconf}
K. Bamba, S. D. Odintsov and D. S\'aez-G\'omez, {\em Conformal symmetry and accelerating cosmology in 
teleparallel gravity}, Phys. Rev. {\bf D 88} (2013) 084042, [1308.5789]. 

\bibitem{MM-TPconf}
D. Momeni and R. Myrzakulov, {\em Conformal Invariant Teleparallel Cosmology}, Eur. Phys. J. Plus {\bf 129} 
(2014) 137, [1404.0778]. 

\bibitem{wri-TPconf}
M. Wright, {\em Conformal transformations in modified teleparallel theories of gravity revisited}, Phys. 
Rev. {\bf D 93} (2016) no.10, 103002, [1602.05764].


\bibitem{RAS-MMT-1}
H. Ramo Chothe, A. Dutta and S. Sur, {\bf in preparation}.

\bibitem{RAS-MMT-2}
S. Sur, A. Dutta and H. Ramo Chothe, {\bf in preparation}.

\bibitem{mal-CG}
J. Maldacena, {\em Einstein Gravity from Conformal Gravity}, 1105.5632. 

   
\bibitem{AMS2000-kess}
C. Armendariz-Picon, V. Mukhanov and P. J. Steinhardt, {\em A Dynamical Solution to the Problem of a 
Small Cosmological Constant and Late-time Cosmic Acceleration}, Phys. Rev. Lett. {\bf 85} (2000) 4438
[astro-ph/0004134].

\bibitem{AMS2001-kess}
C. Armendariz-Picon, V. Mukhanov and P. J. Steinhardt, {\em Essentials of k-essence}, Phys. Rev. 
{\bf D 63} (2001) 103510, [astro-ph/0006373].

\bibitem{MCLT-kess}
M. Malquarti, E. J. Copeland, A. R. Liddle and M. Trodden, {\em A New view of k-essence}, Phys. Rev. 
{\bf D 67} (2003) 123503, [astro-ph/0302279].

\bibitem{schr-kess}
R. J. Scherrer, {\em Purely kinetic k-essence as unified dark matter}, Phys. Rev. Lett. {\bf 93} (2004)
011301, [astro-ph/0402316].

\bibitem{sssd-kess}
S. Sur and S. Das, {\em Multiple kinetic k-essence, phantom barrier crossing and stability}, JCAP 
{\bf 0901} (2009) 007, [0806.4368].


\bibitem{ssasb-tcons}
S. Sur, A. S. Bhatia, {\em Constraining Torsion in Maximally symmetric (sub)spaces}, Class. Quant. Grav. 
{\bf 31} (2014) 025020, [1306.0394].



\end{thebibliography}
\end{document}